\documentclass[a4paper,11pt]{article}
\pdfoutput=1 

\usepackage{jheppub} 

\usepackage[T1]{fontenc} 

\usepackage[T1]{fontenc} 
\usepackage{physics}
\usepackage{amsmath}
\usepackage{dsfont}
\usepackage{graphicx}
\usepackage{bbold}
\graphicspath{ {./1 /} }

\title{
\boldmath
Explicit reconstruction of the entanglement wedge via the Petz map}

\author[a,b,c,1]{Eyoab Bahiru \note{Corresponding author.}}
\author[a,b]{~and~ Niloofar Vardian }

\affiliation[a]{SISSA, International School for Advanced Studies, via Bonomea 265, 34136 Trieste, Italy}
\affiliation[b]{INFN, National Institute for Nuclear Physics, Sezione di Trieste, Via Valerio 2, 34127 Trieste, Italy}
\affiliation[c]{ICTP, International Centre for Theoretical Physics, Via Strada Costiera 11, 34151, Trieste, Italy}

\emailAdd{ebahiru@sissa.it}
   \emailAdd{nvardian@sissa.it}

\abstract{We revisit entanglement wedge reconstruction in AdS/CFT using the Petz recovery channel. In the case of a spherical region on the boundary, we show that the Petz map reproduces the AdS-Rindler HKLL reconstruction.  
Moreover, for a generic subregion of the boundary, we could obtain the same boundary representation of a local bulk field lies in the entanglement wedge as the one proposed earlier in \cite{jafferis2016relative, faulkner2017bulk} using properties of the modular flow.}

\begin{document} 
\maketitle
\flushbottom

\section{Introduction}
\label{sec:intro}

An important question in the AdS/CFT correspondence is that of {\it subregion duality}: is it possible to associate regions of the bulk to specific regions of the boundary CFT? 

Given a spacelike region $A$ in the CFT, one can associate two candidate dual regions in the bulk to it. One is the {\it causal wedge} of region $A$. This is constructed by first considering the boundary domain of dependence $\mathcal{D}(A)$ of $A$ and then considering all bulk points which are both in the causal future and causal past of $\mathcal{D}(A)$. The other one is the {\it entanglement wedge} $ \mathcal{E}(A)$ of the region $A$, defined as the bulk domain of dependence of a bulk spacelike  surface whose boundary is the union of $A$ and the Ryu-Takayanagi surface \cite{Ryu:2006bv} associated to $A$ \footnote{When the bulk geometry is time-dependent, one has to take what is called HRT surface\cite{Hubeny:2007xt}, generalization of Ryu-Takayanagi surface. Later, considering quantum corrections to HRT surfaces led to the conjecture of quantum extremal surfaces\cite{Engelhardt:2014gca, Dong:2017xht}, which should be used at higher orders in perturbation theory.}. 
The entanglement wedge contains the causal wedge, and in general, it is larger than the causal wedge \cite{Headrick:2014cta,Wall:2012uf}. \footnote{The simplest case that it can  happen is when we consider two disconnected regions in the boundary. When such regions are big enough the entanglement wedge is bigger than the causal wedge.}

It is believed that  the entire entanglement wedge of a given boundary region $A$ can be reconstructed from region $A$. This means that bulk operators acting inside the entanglement wedge can be expressed in terms of CFT operators in the region $A$. This idea of \emph{entanglement wedge reconstruction} (EWR) has been introduced and developed in various works \cite{Czech:2012bh, Headrick:2014cta, Wall:2012uf, jafferis2016relative, dong2016reconstruction, cotler2019entanglement}. 
Important evidence in favor of EWR was given in \cite{jafferis2016relative}, where it was argued that the relative entropy of two states in the entanglement wedge is the same as that in the corresponding boundary region.
Using this argument, the authors in \cite{dong2016reconstruction} could prove that for large class of states the bulk region dual to a given region of the boundary is its entanglement wedge\footnote{For subtleties involving the entanglement wedge, see \cite{Hayden:2018khn,Akers:2020pmf}. }.
If EWR holds, then it follows that the causal wedge of $A$ can also be reconstructed from $A$, as it is generally smaller than the entanglement wedge.

Causal wedge reconstruction is well understood at large $N$. Using the bulk equations of motion and the fact that bulk fields asymptote to CFT local operators near the boundary of AdS (also called the extrapolate dictionary), it is possible to express bulk operators in the causal wedge in terms of smeared single-trace operators in the causal domain of the boundary region $A$. 
This approach is called HKLL reconstruction which
has been introduced in 
 seminal of papers
 \cite{Banks:1998dd, Bena:1999jv,hamilton2006local, hamilton2006holographic, hamilton2007local, hamilton2008local}. A simple example of a subregion where this can be worked out explicitly is the case of the AdS-Rindler wedge, where via an HKLL approach one can express bulk operators in the AdS-Rindler subregion in terms of CFT operators on the corresponding boundary subregion.
 It is generally believed that this procedure can be extended to all orders in $1/N$, by adding multi-trace corrections \cite{Kabat:2013wga}.

On the other hand EWR is more subtle. This is especially true in cases where the entanglement wedge is larger than the causal wedge. The arguments mentioned above in support of EWR are somewhat formal and do not provide us with a concrete representation of bulk operators lie in the entanglement wedge in terms of CFT operators.

A general way to approach the problem of EWR is via the Petz map \cite{petz1986sufficient, petz1988sufficiency}: if a quantum channel between two Hilbert spaces preserves relative entropy then it can be reversed. 
In situations when the relative entropy is only approximately preserved, there is no exact reversibility but one can also use the twirled Petz map \cite{cotler2019entanglement} as an approximation. In our case, the quantum channel is the map from the entanglement wedge of a given boundary region to  the region itself. 
As was argued in \cite{cotler2019entanglement}, starting with an isometry that  maps the entire bulk Hilbert space to the entire boundary Hilbert space, such as the one related to the global HKLL reconstruction,
one can find the explicit form of the quantum channel we are interested in.
Then, the dual of the corresponding Petz recovery channel, called Petz map, allows us - in principle - to express operators in the entanglement wedge in terms of operators in the region $A$.
Moreover, considering $1/N$ corrections,  it was argued in \cite{chen2020entanglement} that the Petz map is still good enough in finite-dimensional code subspaces whose dimensions do not grow exponentially in $N$, as long as the error is non perturbatively small. 

However, the expression resulting from the Petz map for the bulk operator in terms of boundary data is still somewhat abstract. The relevant formulae involve a projector on the code subspace and taking a trace over part of the CFT which in practice, it is not easy to treat. Until now, there are no examples where the Petz map expression has been computed in detail in terms of simple CFT quantities. 

In this paper, we make two advances towards a better understanding of the EWR via the Petz map. First, we demonstrate how, when applied to an AdS-Rindler wedge and working at large $N$, the Petz formula  reproduces the standard  AdS-Rindler HKLL reconstruction. While this has been generally assumed to be true, to our knowledge it has not been explicitly demonstrated\footnote{In \cite{cotler2019entanglement}, the authors used the twirled Petz map (check subsection \ref{sec.2.4}) to reconstruct AdS-Rindler wedge in a very restrictive case when the code subspace contains only the vacuum and one particle state.}. Second, we consider more general subregions, where the entanglement wedge may even be larger than the causal wedge, and we apply the Petz formula and show that  it reproduces results on EWR which were earlier conjectured in \cite{jafferis2016relative} and also derived in \cite{faulkner2017bulk} using arguments based on modular flow.
The crucial point in our work is using the Reeh-Schlieder theorem in QFT.  Based on this theorem, we can define the code subspace by acting just with operator algebra of the specified subregion on the corresponding semi-classical state. 

The plan of the paper is as follows: in section 2, we review some basic aspects of bulk reconstruction using the HKLL approach. In section 3, we introduce the concept of a quantum channel, we discuss conditions for its reversibility and introduce the Petz map. In section 4, we review the proposal of \cite{cotler2019entanglement}, of how the Petz map can be used for EWR. In section 5, we show how in the case of the AdS-Rindler wedge the Petz map reproduces the more standard HKLL AdS-Rindler reconstruction. In section 6, we apply the Petz map to more general entanglement wedges.

\section{Bulk reconstruction in AdS/CFT }

According to the AdS/CFT correspondence, a holographic CFT on $ {\mathbb R}\times {\mathbb S}^{d-1}$
can be interpreted as a theory of quantum gravity in an asymptotically $AdS_{d+1} \times M$ spacetime, where $M$ is some compact manifold. Usually, this involves taking a large $N$ limit in the CFT and bulk interactions are suppressed by powers of $1/N$. Thus, to leading order at large $N$, the bulk quantum theory consists of free fields.

The correspondence also involves an identification between fields in the bulk and operators in the boundary CFT. For example, the CFT operator dual to a bulk scalar field $\phi$ is a scalar primary $O$ with conformal dimension $ \Delta$ related to the mass of the field $\phi$ by $\Delta = d/2 + \sqrt{ m^2 + d^2/4} $ and the extrapolate dictionary defines $O$ as the dual of $\phi$ at infinity. 
For simplicity, in the following, we will just focus on scalar fields and discuss the identification with the dual CFT operator $O$ at large $N$.

First, on the bulk side of the duality, we start with AdS$_{d+1}$ in global coordinates
$(t , \rho, \Omega)$ which is described with the  metric below
\begin{equation}
    ds^2 = \frac{1}{\cos^2(\rho)} 
    \big(- dt^2 + d\rho ^2 + \sin^2(\rho) d\Omega ^2 _{d-1}\big).
\end{equation}
Consider a scalar field on the $ AdS_{d+1}$ background with the action
\begin{equation}
    S= \int d^{d+1}x ~\sqrt{-g}~ \frac{1}{2}~ \big(~ g^{\mu\nu} \nabla_{\mu} \phi \nabla_{\nu} \phi - m^2 \phi ^2 \big)
    \end{equation}
and corresponding equation of motion 
\begin{equation}
\label{kgequation}
(\Box - m^2)\phi=0.
\end{equation}
This equation has to be supplemented with normalizable boundary conditions at infinity, which implies that near the AdS boundary $\rho={\pi \over 2}$, the field has to decay as $\phi \sim (\cos\rho)^{\Delta}$. With these boundary conditions at infinity and demanding regularity in the interior, we find a basis of solutions for \eqref{kgequation} denoted as $ f_{nlm}(t, \rho, \Omega)$ which is labeled by the quantum numbers $n,l$ and $m$, where $ n \in \{0,1,2,...\}$, $l$ is the total angular momentum of the corresponding mode and $m$ is related to the other angular quantum numbers needed to specify a mode.
These modes  are proportional to 
\begin{equation}\label{21}
    f_{nlm}(t, \rho, \Omega) \propto e^{-i E_{nl}t} Y_{lm}(\Omega) \sin^l (\rho) \cos^\Delta (\rho)
    P_n ^{(l+d/2-1, \Delta-d/2)} (\cos 2\rho)
\end{equation}
while
\begin{equation}\label{22}
     E_{nl} = \Delta + 2n + l,
\end{equation}
and $\Delta=d/2 + \sqrt{m^2+d^2/4}$ is the conformal dimension of the dual CFT$_d$ operator $O$.

To quantize the scalar field, we associate an annihilation operator $ a_{nlm}$ to each mode $ f_{nlm}$ with normalized commutation relation
\begin{equation}
    [a_{nlm}, a^\dagger _{n'l'm'}] = \delta _{nn'} \delta _{ll'} \delta _{mm'}.
\end{equation}
The quantized free scalar field in AdS$_{d+1}$ is given by 
\begin{equation}\label{60}
    \phi (t, \rho, \Omega) = \sum _{nlm} f_{nlm} (t, \rho, \Omega) a_{nlm} + f^* _{nlm} (t, \rho, \Omega) a^\dagger _{nlm}.
\end{equation}
The modes $ f_{nlm}(t, \rho, \Omega)$ should be normalized in such a way that the field $\phi$ obeys the canonical commutation relation. To find the correct normalization factor we consider the Klein-Gordon inner product defined on a Cauchy surface $ \Sigma$. 
If we assume that $ t$ direction is orthogonal to $ \Sigma$,  for every two functions $\phi_1$ and $ \phi_2$, it is defined as
\begin{equation}
    \langle\phi_1 , \phi_2\rangle _{KG} \equiv i \int _\Sigma d^d x ~\sqrt{-g}~ g^{tt} \big(\phi_1 ^* \nabla_t \phi _2 -  \phi_2\nabla_t \phi _1 ^*\big).
\end{equation}
If both  $\phi_1$ and $ \phi_2$ obey the equation of motion, the integral above defines a conserved inner product in $t$. In particular, it says that if we normalize the modes $f_{nlm}$ at some time such that 
$ \langle f_{nlm} , f_{n'l'm'}\rangle = \delta _{nn'} \delta _{ll'} \delta _{mm'} $ and $ \langle f_{nlm} , f^\dagger_{n'l'm'}\rangle  = 0$, they will remain normalized also at later times \cite{kaplan2016lectures}.
Following these steps, in the end, one can write the modes explicitly as
\begin{equation}\label{30}
 f_{nlm}(t, \rho, \Omega) = \frac{1}{N_{nlm}} e^{-i(\Delta + 2n + l)t} Y_{lm}(\Omega) \sin^l (\rho) \cos^\Delta (\rho)   P_n ^{(l+d/2-1, \Delta-d/2)} (\cos 2\rho)   
\end{equation}
where
\begin{equation}
    N_{nlm} =\sqrt{ \frac{\Gamma(n+l+d/2)\Gamma(n+\Delta-d/2 +1)}{n! \Gamma(n+l+\Delta)}}.
\end{equation}

The conformal boundary of AdS$_{d+1}$ is the cylinder $ {\mathbb R} \times {\mathbb S}^{d-1}$ which in terms of the global coordinates we obtain by taking $ \rho \rightarrow \pi /2$ limit. We can use the coordinate $t$ and $\Omega$ to parametrize the boundary theory  with metric
\begin{equation}
     ds^2 = - dt^2 +  d\Omega ^2 _{d-1}.
\end{equation}
In the boundary, using the state-operator correspondence in the CFT, the formula (\ref{22}) has a nice interpretation. The state created by the $n=l=0$ creation operator is identified with the state in the CFT that is produced by inserting the single-trace primary operator $O$ dual to $ \phi$ into the center of the Euclidean path integral and other excited states come from inserting its descendants. 

More generally, to leading order at large $N$, the Fourier modes $ O_{nlm}$ of the single trace primary operator and $ a_{nlm}$ for the mode $ f_{nlm}$ are the same up to apriori arbitrary constant $M_{nlm}$.
The extrapolate dictionary in the global coordinates is given by
\begin{equation}\label{25}
    O(t,\Omega) = \lim _{\rho \rightarrow \pi/2} \frac{1}{\cos ^ \Delta \rho } ~\phi (t, \rho, \Omega). 
\end{equation}
As a result, we can define the CFT operator $ \hat{O}_{nlm} = \frac{1}{M_{nlm}} O_{nlm}$
which is identified with the bulk operator
\begin{equation}\label{55}
    \hat{O}_{nlm} = a_{nlm}.
\end{equation}
As we will see later, this allows us to write a CFT expression for a local bulk field at any point in the bulk.

The single trace primary operator $O$ has a mode expansion on $ {\mathbb R} \times {\mathbb S}^{d-1}$ as
\begin{equation}\label{46}
    O (t, \Omega)=   \sum _{nlm} g_{nlm} (t,\Omega) O_{nlm} + g^* _{nlm} (t, \Omega) O^\dagger _{nlm}.
\end{equation}
Following \ref{60}, we have $ g_{nlm} = \frac{1}{M_{nlm}} \lim _{\rho \rightarrow \pi/2} \frac{1}{\cos ^ \Delta \rho }~ f_{nlm}(t, \rho, \Omega)$. Thus $M_{nlm}$ can be chosen so that mode functions $g_{nlm}$ are orthonormal. 

At the large $N$ limit, since we have a free theory in the bulk, all correlators can be reduced to products of 2-point functions by Wick contractions. Therefore, on the boundary side, we already know all the n-point functions of the operator $ O$ by taking the spacetime points to the boundary in the expression we found for the bulk and using the extrapolate dictionary (\ref{25}), we have
\begin{align}  \label{23}
    \langle O(x_1)  O(x_2) \rangle &\propto \frac{1}{(x_1 - x_2)^{2\Delta}}\\  \label{24}
    \langle O(x_1)  O(x_2) ...  O(x_{2n}) \rangle  & = \langle O(x_1)  O(x_2) \rangle ... \langle O(x_{2n-1})  O(x_{2n}) \rangle + permutations.
\end{align}
Although the correlation functions of $O$ factorize to the product of 2-point functions, the scalar primary operator is not really a free scalar field. In a CFT in $d$ spacetime dimensions, the condition that a scalar operator is free, i.e. $ \nabla^2 O =0 $, is equivalent to the fact that its conformal dimension is $ \Delta = d/2 -1$. Therefore, as the conformal dimension for the scalar primary operator $O$ in a holographic CFT is $\Delta = d/2 + \sqrt{ m^2 + d^2/4} $, it is actually not a free scalar theory on the boundary.
For the free scalar primaries, the factorization is a consequence of the equation of motion. More generally, the scalar fields with $ \Delta\geq d/2-1 $ are called \emph{generalized free fields} (GFFs) \cite{GREENBERG1961158,Duetsch:2002hc,el2012emergent} if their correlators take the form of Eqs. (\ref{23}) and (\ref{24}). However, because they do not obey the linear equation of motion, we can not describe them in terms of a local free lagrangian in the spacetime background in which the CFT lives.
The reason that such fields can be called free is that their Hilbert space has a Fock space structure that is the Hilbert space of a free theory. Nevertheless, such a CFT, extrapolated to high temperatures,  has the wrong thermodynamic properties, and therefore it is inconsistent by itself. For the operators with large conformal dimensions, the spectrum can not have the structure of a freely generated Fock space. One way to solve the problem is that think about the GFF as the low-conformal dimension sector of a much larger CFT with a large central charge while all the additional states correspond to the black hole microstates in the bulk \cite{el2012emergent}.

As a result, we observe that at the large $N$ limit, a free scalar field in pure AdS can be identified as  GFF of the boundary.

\subsection{HKLL reconsruction method: mode sum approach}\label{modesum}

The study of bulk reconstruction, usually called HKLL, was developed by
 Hamilton, Kabat, Lifschytz and Lowe
in a series of papers \cite{Banks:1998dd, Bena:1999jv,hamilton2006local, hamilton2006holographic, hamilton2007local, hamilton2008local}. 
They attempt to reconstruct the operators of the
bulk gravitational theory in the non-interacting regime from the operators of the boundary.
Bulk operators are expressed in terms of smeared single trace operators in the CFT as
\begin{equation}
    \phi (X) \quad \longleftrightarrow \quad \int dx~ K(X|x) O(x),
\end{equation}
where the kernel $K(X|x) $ is called \emph{smearing function}. At large $N$ limit, finding the smearing function can be implemented through Fourier expansion. Consider $ f_n(X)$ as the set of orthogonal solutions to the Klein-Gordon equation. For simplicity here we denote  the set of labels $(nlm)$ discussed in the previous subsection collectively by $n$. One can quantize the bulk field in terms of creation and annihilation operators as
\begin{equation}\label{32}
    \phi(X)= \sum_n f_n(X) a_n +h.c.
\end{equation}
Taking the points to the boundary and using the extrapolate dictionary, we get the mode expansion of the boundary operators as 
\begin{equation}\label{31}
    O(x)= \sum_n \tilde{g}_n(x) a_n +h.c.
\end{equation}
where $ \tilde{g}_n(x) = \lim_{r\rightarrow \infty} r^\Delta f_n(r,x)$. If one defines orthonormal boundary mode functions $g_n(x)=\frac{1}{M_{n}} ~\tilde{g}_n(x)$, one can invert (\ref{31}) and obtain
 \begin{equation}
   a_n = \frac{1}{M_{n}}\int dx O(x) g_n^*(x) .
\end{equation}
By plugging it back to  (\ref{32}), we reach
\begin{equation}\label{33}
    \phi(X)= \sum_n \frac{1}{M_{n}} f_n(X)\int dx ~O(x) \tilde{g}_n^*(x) +h.c..
\end{equation}
In case we are able to exchange the summation and integration in (\ref{33}), we will get 
\begin{equation}
    \phi (X)=\int dx ~K(X|x) O(x), 
\end{equation}
where
$  K(X|x) = \sum_n M_{n} ^{-1} f_n(X)~\tilde{g}_n^*(x) + h.c. $ is the smearing function \cite{leichenauer2013ads}.
We review the HKLL construction for a free scalar field in pure AdS in global and  AdS-Rindler coordinates in appendix \ref{A}.

\subsection{Bulk reconstruction and subregion duality }\label{sec. Heisenberg picture}

As we had in the previous sections, a bulk operator $ \phi(X)$ can be represented as a smearing integral of boundary operators
\begin{equation}
    \phi(X)= \int _{bdy} d^{d-1}x dt~ K(X|t,x) O(t,x) + O(1/N).
\end{equation}
We can use the CFT Hamiltonian to re-express all operators $ O(t,x)$ in terms of fields on a Cauchy surface $\Sigma$ of the boundary by explicitly evolving the operators with the boundary Hamiltonian. Let us consider the pure AdS case and $\Sigma$ to be the $t=0$ slice
\begin{equation}\label{36}
    \phi(X)= \int _{bdy} d^{d-1}x ~dt ~K(X|t,x)~ e^{iH_{CFT}t} O(x) e^{-iH_{CFT}t}
\end{equation}
where $ O(t=0,x) = O(x)$. In general, operators of the form $e^{iH_{CFT}t}~ O(x)~ e^{-iH_{CFT}t}$ have support on a large part of the slice $\Sigma,~ t=0$. 
An interesting question in AdS/CFT  is whether the CFT representation of $ \phi(X)$ can be localized to a subregion of $\Sigma$. Intuitively, it is expected that the boundary support of $\phi(X)$ shrinks as the operator approaches the boundary. 
However, one can see from (\ref{36}) that even if we take $X$ to be very close to the boundary, the CFT reconstruction still has support on the entire $\Sigma$. 

In fact, it is possible to reconstruct bulk operators so that they are supported on smaller regions on the boundary. Consider a spherical subregion $R$ on $\Sigma$ and the corresponding causal wedge in the bulk. Consider a local field $\phi(X)$ localized inside this causal wedge. Then it is possible to represent the bulk field as
\begin{equation}
    \phi(X)= \int _{\mathcal{D}(R)}  d\tau  d^{d-1}x ~K_R(X|\tau,x) O(\tau,x).
\end{equation}
where the integral is over the domain of dependence ${\cal D}(R)$ of $R$ and $K_R(X|\tau,x)$ is a new smearing function called the AdS-Rindler smearing function.

Again we can write it in terms of non-local operators in the Heisenberg picture which evolves with Rindler Hamiltonian $ H_{\tau}$
\begin{equation}\label{37}
    \phi(X)=  \int_{-\infty}^{\infty}  d\tau \int _{R} d^{d-1}x ~K_R(X|x,\tau) ~e^{iH_{\tau}\tau} O(x) e^{-iH_{\tau}\tau}.
\end{equation}
The operators $e^{iH_{\tau}\tau}~ O(x)~ e^{-iH_{\tau}\tau}$ are again some non-local operators but this time they have support only on region $R$ instead of entire $ \Sigma$.
Therefore, the AdS-Rindler reconstruction provides us a way to localize the representation of $\phi(X)$ in the boundary. More generally, it suggests the proposal that a given region $R$ on a Cauchy slice of the boundary encodes the bulk data inside the causal wedge of its boundary domain of dependence. 

Nevertheless,
one can go ahead and look at the Rindler Hamiltonian in \eqref{37} as the 
\emph{modular Hamiltonian} of the region $R$ that generates the modular flow of operators on $R$. For the case of AdS-Rindler, it is  just translation in the $\tau$ direction. In \cite{jafferis2016relative}, authors showed that the boundary modular flow is dual to the bulk modular flow in the entanglement wedge $\mathcal{E}_R$ and conjectured that operators in the entanglement wedge of the region $R$ are the ones can be constructed on the boundary region $R$ by replacing $\tau$ in (\ref{37}) by the modular parameter $s$  as
\begin{equation}\label{38}
    \phi(X)= \int _{R} d^{d-1}x \int_{-\infty}^{\infty} ds~ K'_R(X|x,s) O(x,s) 
\end{equation}
for every $ X \in \mathcal{E}_R $ where $ O(x,s)= e^{iK_R s} ~O(x, s=0) ~e^{-iK_R s}$.

\section{Background on quantum channels}

In this section, we introduce the notion of a {\it quantum channel} and the conditions under which it is reversible. The central point of this section is the introduction of the \emph{Petz map} \eqref{14}, which will be used in later sections in the context of bulk reconstruction.

\subsection{Quantum channels}

Real systems suffer from unwanted interactions with the outside world that show up as noise in quantum information processing systems. In order to describe such systems, it is useful to introduce the notion of a quantum channel, i.e. a linear map $\mathcal{E}: L({\cal H}) \rightarrow L({\cal H})$ which is completely positive and trace-preserving. 
Here, $L({\cal H})$ denotes the set of linear operators acting on the Hilbert space ${\cal H}$.\footnote{More generally a quantum channel can be a map between $L({\cal H}_1), L({\cal H}_2)$ for two different Hilbert spaces ${\cal H}_1,{\cal H}_2$.} Every quantum channel 
$\mathcal{E} $ has an operator sum representation in terms of a non-unique set of operators $ \{ A_ i\}$ known as \emph{Kraus operators} \cite{1983LNP...190.....K,hellwig1970operations,preskill1998lecture} such that,
 \begin{equation}
      \mathcal{E} ( \rho ) =  \sum _i A_i  \rho A_i^ \dagger  \qquad\qquad  \sum _i A_i^ \dagger A_i = I.
 \end{equation}
They are the most general transformation of a quantum state in an open quantum system.

 A natural way to describe the dynamics of an open quantum system is to regard it as arising from an interaction between the system and an environment which together transform under a unitary. After the evolution
 we perform a partial trace over the environment to obtain the state of the system. 
 For every quantum channel, there exists a model environment starting in an initial state $ \sigma_{en} $ and model dynamics specified by a unitary operator $ U $ such that 
 \begin{equation}\label{1}
     \mathcal{E} (.) = \tr_{en} \big( U(. \otimes \sigma _{en} ) U^\dagger \big),
 \end{equation}
which is a version of the \emph{Stinespring dilation}  theorem.
 If $\sigma_{en}  = \sum _j \lambda_j \ket{j}\bra{j} $, one can find the Kraus representation of (\ref{1}) as
\begin{equation}
    \mathcal{E} ( \rho ) =  \sum _{j,k} A_{j,k} ~\rho~ A_{j,k}^\dagger
\end{equation}
which $A_{j,k} = \sqrt{\lambda _j} \bra{k}U\ket{j}   $
are the Kraus operators. Therefore, we can describe the dynamics of the system by using the operator-sum representation without having to explicitly consider the properties of the environment.
 One advantage of this Kraus representation is that it is well adapted to describe discrete state change without explicit reference to the passage of time.

\subsection{Universal recovery channel and the Petz map}\label{sec.2.4}

A quantum channel $\mathcal{E}$ is called reversible if one can find a \emph{recovery channel} $\mathcal{R}:  L({\cal H}) \rightarrow L({\cal H})$ such that
\begin{equation}\label{5}
    \mathcal{R} \circ \mathcal{E} (\rho)=\rho
    \qquad \forall \rho \in S({\cal H}).
 \end{equation}
Most quantum channels, which correspond to open or noisy systems, are not reversible.
We will return to the question of reversibility later in this subsection.

 In order to do quantum information and communication in the presence of noise, we need \emph{quantum error correction} (QEC) codes. 
The basic ideas of the theory are inspired by  classical information theory, but all the limitations of quantum mechanics have been considered in its formulation. 
A quantum error correcting code corresponds to selecting an appropriate subspace, called \emph{code subspace}($\mathcal{C}$ or $\mathcal{H}_{code} $) that has the same dimension as the system, of some larger Hilbert space. In the general theory of QEC, the noise model is described by a quantum channel $\mathcal{E}$. 
The code subspace can be corrected if we can find a recovery channel $\mathcal{R}$, such that for every state $\rho$ whose support lies  within $\mathcal{H}_{code}$,
the channel can be reversed, i.e.
\begin{equation}
    \mathcal{R} \circ \mathcal{E} (\rho)=\rho
    \qquad \forall \rho = P \rho P
 \end{equation}
where $P$ is the projection into the code subspace.
One might be interested to consider a physical system instead of a code subspace. In such a case, if we take 
$V: \mathcal{H}_{system} \rightarrow \mathcal{H}$ as the isometry that embeds the $\mathcal{H}_{system}$ into $\mathcal{H}$, we can rewrite (\ref{5}) as the following
\begin{equation}
\mathcal{R} \circ \mathcal{E} (V\rho V^\dagger)=V\rho V^\dagger
    \qquad \forall \rho \in S(\mathcal{H}_{system})
\end{equation}
that is equivalent to having
$ \mathcal{E}'$
 and $\mathcal{R}'$ such that 
$\mathcal{R}' \circ \mathcal{E}' (\rho )=\rho $ where 
$  \mathcal{E}'(.) = \mathcal{E}(V(.)V^\dagger)  $ and
$ \mathcal{R}'(.) =V^\dagger \mathcal{R}(.) V $
\cite{beny2009quantum}.

Given a quantum channel $\mathcal{E}$, it is useful to consider the Hilbert-Schmidt dual channel which defines a mapping of {\it observables} rather than of {\it states}. This is also sometimes referred to as the Heisenberg picture of the channel. The idea is to think of $\mathcal{E}$ as a (discrete) evolution of a state. After the evolution, the result of a measurement of an observable $O$ will be in the form of $ \tr(\mathcal{E}(\rho) O)$, where $\rho$ describes the state of the system. As we usually do when going to the Heisenberg picture, we can alternatively formulate the evolution of the system by transforming the operators, requiring to get the same measurement results. For this purpose, we describe 
the evolution of the observables by the     
channel $\mathcal{E}^* $ that is called \emph{Hilbert-Schmidt dual map} defined as
\begin{equation}
    \tr(\rho\mathcal{E}^* (O) )=\tr(\mathcal{E}(\rho) O) \qquad \forall \rho, O.
\end{equation}
The set of Kraus operators for $\mathcal{E}^*$ is given easily by cyclicity property of trace as
$ \{A_a ^\dagger\}$ instead of $\{A_a\}$, and trace preservation of $\mathcal{E}$ is equivalent to the requirement that $\mathcal{E}^*$ is unital, $\mathcal{E}^* (I)=I$. In the case of QEC, the conservation of a state  by 
$\mathcal{R} \circ \mathcal{E}$ \eqref{5} implies that in the Heisenberg picture for all  the operators $O \in \mathcal{L}(\mathcal{H})$ we have
\begin{equation}
    P( \mathcal{R} \circ \mathcal{E})^* (O) P= P \mathcal{E}^* \circ \mathcal{R}^*(O) P= P O P.
\end{equation}
 



We now return to the general question of the reversibility of a quantum channel. This has been studied widely in \cite{jenvcova2006sufficiency, mosonyi2004structure, petz1986sufficient, petz1988sufficiency}. The reversibility of $\mathcal{E}$ is related to the quantum relative entropy of states under the action of  $\mathcal{E}$. The relative entropy between two states $\rho$ and $\sigma$  is defined as $ S(\rho || \sigma) = \tr(\rho \log \rho - \rho \log \sigma )$ and it is a measure of distinguishability between two quantum states. 
The most important theorem related to this quantity
known as \emph{monotonicity of relative entropy} or the \emph{data processing inequality}, whose proof is discussed in appendix \ref{appendixpetz} for finite dimensions, states that $ S(\rho || \sigma)$ is non-increasing under the action of any quantum channel $\mathcal{E}$ \cite{lindblad1975completely, uhlmann1977relative}, i.e.,
\begin{equation}
\label{monotonicity}
    S(\rho || \sigma) \geqslant S ( \mathcal{E}(\rho)|| \mathcal{E}(\sigma)).
\end{equation}
It has been shown in \cite{petz2003monotonicity,hayden2004structure} that there exists a quantum channel $\mathcal{R}$ such that for all states $\rho \in S({\cal H})$, $  \mathcal{R} \circ \mathcal{E} (\rho)=\rho$  if and only if 
$S(\rho || \sigma) = S ( \mathcal{E}(\rho)|| \mathcal{E}(\sigma))$
for all $ \rho , \sigma \in S({\cal H})$.
Moreover, the explicit form of the quantum channel $\mathcal{R}$ for the set of states 
$\{\mathcal{E}(\rho) | \forall \rho \in S({\cal H})\}$ has been found in \cite{hayden2004structure}. It is given as a function of a reference quantum state 
$\sigma \in S(\mathcal{H}_A)$ and the channel $\mathcal{E}$ itself as
\begin{equation}\label{14}
    \mathcal{R}(.)= \mathcal{P}_{\sigma, \mathcal{E}}(.)= \sigma^{1/2} \mathcal{E}^* \big(\mathcal{E}(\sigma)^{-1/2} (.)\mathcal{E}(\sigma)^{-1/2} \big)\sigma^{1/2}
\end{equation}
where $\mathcal{E}^*$ is the dual channel of $\mathcal{E}$.
$\mathcal{P}_{\sigma, \mathcal{E}}$ is known as \emph{Petz recovery channel}. This result has been also independently obtained by Barnum and Knill in \cite{barnum2002reversing}. We review a proof of the theorem in the finite-dimensional case in appendix \ref{appendixpetz}.

Practically, in most cases, inequality \eqref{monotonicity} will not be saturated hence exact reversibility will not be satisfied. One can ask if there exists an approximate recovery channel that the recovered state be just close to the state $\rho $, 
$ |\mathcal{R}_ \epsilon \circ \mathcal{E} (\rho) - \rho| < \epsilon   $ where $ \mathcal{R}_ \epsilon  $ is approximate version of recovery channel.
In \cite{wilde2015recoverability, sutter2016strengthened}, with two different approaches, it was shown that for any two states $\rho$ and $\sigma$ and channel $\mathcal{E}$ there exists a recovery channel $\mathcal{R}$ such that 
$ \mathcal{R}\circ \mathcal{E} (\sigma)=\sigma $ and
\begin{equation}\label{3}
     S(\rho || \sigma) - S ( \mathcal{E}(\rho)|| \mathcal{E}(\sigma))  \geqslant -2 \log F (\rho , \mathcal{R}\circ \mathcal{E} (\rho))
\end{equation}
where 
$ F(\rho , \sigma) :=  \lVert  \sqrt{\rho} \sqrt{\sigma}\rVert _1 $
is the fidelity of $\rho$ and $\sigma$ that measure the closeness of two quantum states. $F(\rho , \sigma) =1  $ if and only if $\rho = \sigma$, then the inequality of \eqref{3} will be saturated just in the case of exact reversibility. 
In \cite{junge2018universal}, it was shown that the recovery channel $\mathcal{R}$ that satisfies \eqref{3} is universal, which means we can always choose a $\rho$-independent recovery channel.
Furthermore, they could find the explicit expression for the universal recovery map $ \mathcal{R}_{\sigma , \mathcal{E}}$ as
\begin{equation}\label{4}
    \mathcal{R}_{\sigma , \mathcal{E}}(.)=
    \int _\mathbb{R} dt~ \beta_0 (t)~ 
    \sigma ^{-it/2}
    \mathcal{P}_{\sigma, \mathcal{E}}\big(\mathcal{E}(\sigma)^{it/2} (.)\mathcal{E}(\sigma)^{-it/2}\big)
    \sigma ^{-it/2}
\end{equation}
that is called \emph{twirled Petz map} where $\mathcal{P}_{\sigma , \mathcal{E}}$
is the Petz recovery channel given in \eqref{14} and $ \beta_0 (t) = \frac{\pi}{2} ( \cosh(\pi t) +1 )^{-1}$. 
In the case of exact reversibility, the expression \eqref{4} is reduced to the Petz recovery channel.

\section{Entanglement wedge reconstruction via universal recovery channels}

In this section, we review the arugments of \cite{cotler2019entanglement}, on how the Petz map can be used to reconstruct bulk operators in the {\it entanglement wedge} of a boundary subregion.

 \subsection{Background}

Before we proceed we introduce an ingredient that will be useful in what follows. This is the idea of a {\it code subspace} around a given state. For example, starting with the global AdS vacuum state $|\Omega\rangle$ we define
\begin{equation}\label{9}
    \mathcal{H}_ \mathcal{C} = span \{ \ket{\Omega}, \phi _i (x) \ket{\Omega}, ... ,\phi _i (x_1)\phi _j (x_2) \ket{\Omega} , ...\},
\end{equation}
where the range of $i$ and the number of $\phi$ insertions are finite. More generally we can define the code subspace around any semi-classical state. 
This subspace is the one where low-energy experiments in the bulk can be described and we will study bulk reconstruction within a given code subspace.

The entanglement wedge of a boundary region $A$ is defined as the bulk domain of dependence of any bulk spacelike  surface whose boundary is the union of $A$ and the codimension two extremal area surface of minimal area (more precisely, quantum extremal surface) whose boundary is $ \partial A$. It is generally believed that bulk operators inside the entanglement wedge can be reconstructed by operators in the region $A$ on the boundary. 

An important ingredient supporting this is the observation of JLMS \cite{jafferis2016relative} that the relative entropy of two states in the boundary region $A$ is equal to the relative entropy of the two corresponding bulk states in $\mathcal{E}_A $ up to subleading correction. 
\begin{equation}
\label{jlmseq}
    S(\rho_ A|| \sigma _A )=S(\rho_ a|| \sigma _ a) + O(1/N)
\end{equation}
which already suggests that information in the entanglement wedge is contained in region $A$ on the boundary. Using \eqref{jlmseq} arguments in favor of entanglement wedge reconstruction were given in \cite{dong2016reconstruction}.

 Assume that the bulk Hilbert space has a decomposition as $\mathcal{H}_{bulk} = \mathcal{H}_a \otimes \mathcal{H}_{\Bar{a}} $, while $ a= \mathcal{E}_A$. For the cases where the setup is symmetric, like the vacuum sector of the system, the complement region of $a$ in the bulk is also the entanglement wedge of the region $\Bar{A}$ in the boundary, $ \Bar{a}= \mathcal{E}_ {\Bar{A}}$, so the same argument applies for $ \Bar{A}$ and $ \Bar{a}$.
 In general, the entanglement wedge of a given boundary region $A$ can be bigger than its causal wedge.
 Finally, 
 the entanglement wedge reconstruction is a statement that says  any bulk operator $\phi_a$ acting within $\mathcal{H}_a$ can always be represented in the CFT as an operator $O_A$ has support only on $\mathcal{H}_A$.

 \subsection{Entanglement wedge reconstruction with a universal recovery channel}

We now discuss entanglement wedge reconstruction in terms of the universal recovery channels described in Sec. \ref{sec.2.4}, based on \cite{cotler2019entanglement}.

First, consider the entanglement wedge reconstruction and for simplicity assume both bulk and CFT Hilbert spaces have a tensor decomposition as 
$\mathcal{H}_{bulk} = \mathcal{H}_a \otimes \mathcal{H}_{\Bar{a}}~ $
and
$~\mathcal{H}_{CFT} = \mathcal{H}_A \otimes \mathcal{H}_{\Bar{A}} $
.  At large $N$ when the equality between the relative entropy of the states in the entanglement wedge and the boundary region $A$ is exact, i.e.,
 \begin{equation}\label{13}
     S(\rho_ A|| \sigma _A )=S(\rho_ a|| \sigma _ a)
 \end{equation}
from the discussion in Sec. \ref{sec.2.4}, one can say that there exists a quantum channel $\mathcal{R}$ which 
recovers the information in the entanglement wedge from the boundary region $A$. Using the dual channel $\mathcal{R}^*$
we can map operators on $\mathcal{H}_a$ to operators on $\mathcal{H}_A$ as 
$ O_A = \mathcal{R} ^* (\phi _a )$.

If we assume that there is no black hole in the bulk, the global HKLL reconstruction reviewed in section 2 provides us a map from states of the entire bulk to states of the entire boundary.  We can therefore define an isometry of embedding  $ V_{HKLL}$ that embeds the bulk effective field theory Hilbert space to the CFT Hilbert space
$ V_{HKLL} : \mathcal{H}_{bulk} \hookrightarrow \mathcal{H}_{CFT}$,
which
$ \mathcal{H}_{code} = V_{HKLL}~ \mathcal{H}_{bulk}~ V^\dagger _{HKLL}$. 

We now define a quantum channel $\mathcal{E}: S(\mathcal{H}_a) \rightarrow S(\mathcal{H}_A)$. Here $S(\mathcal{H}_a)$ denotes the set of possible density matrices in the bulk region $a$ while $S(\mathcal{H}_A)$ is the set of density matrices in the boundary region $A$.
As the entire AdS space is a closed system, the noise model $\mathcal{E}: S(\mathcal{H}_a) \rightarrow S(\mathcal{H}_A)$ can be written in terms of a model environment (\ref{1}) using the global HKLL map.
We take the complementary bulk region $ \Bar{a}$ in a fixed reference state $ \sigma _{\Bar{a}}$ and then, we can write the quantum channel $ \mathcal{E}$ as
\begin{equation}\label{16}
  \mathcal{E} (.) = \tr_{\Bar{A}} \big( V_{HKLL}(. \otimes \sigma _{\Bar{a}} ) V_{HKLL}^\dagger\big).
\end{equation}
 To map the operators, as we had in Sec. \ref{sec.2.4}, one can go to the Heisenberg picture and write the dual of Petz recovery channel of $\mathcal{E}$ by taking a fixed reduced density matrix on the entanglement wedge $ \sigma_a$,  using expression \eqref{14}, we reach to
\begin{equation}
    O_A = \mathcal{R} ^* (\phi _a )
    =\mathcal{E}(\sigma _a)^{-1/2}
     \mathcal{E}  \big(\sigma _a^{1/2} \phi _a \sigma_a^{1/2}\big)\mathcal{E}(\sigma _a)^{-1/2},
\end{equation}
which for the quantum channel \eqref{16}, it will give us
\begin{equation}\label{17}
    O_A
    =\mathcal{E}(\sigma _a)^{-1/2}
     \tr_{\Bar{A}} \big(V_{HKLL}(\sigma _a^{1/2} \otimes \sigma _{\Bar{a}}^{1/2}) (\phi _a \otimes I_{\Bar{a}}) (\sigma _a^{1/2} \otimes \sigma _{\Bar{a}}^{1/2}) V_{HKLL}^\dagger\big)
     \mathcal{E}(\sigma_a)^{-1/2},
\end{equation}
where
$\mathcal{E}(\sigma_a)= \tr_{\Bar{A}} \big( V_{HKLL}(\sigma_a \otimes \sigma _{\Bar{a}} ) V_{HKLL}^\dagger\big)$.
If we take both $ \sigma_a$ and $ \sigma_{\Bar{a}}$ two maximally mixed states or equivalently putting the bulk in the maximally mixed state $ \tau$, the map (\ref{17}) will be simplified as 
\begin{equation}\label{18}
    O_A
    =\frac{1}{d_{code}}
    \tau _A ^{-1/2}
     \tr_{\Bar{A}} \big(V_{HKLL}(\phi _a) V_{HKLL}^\dagger\big)
  \tau _A^{-1/2},
\end{equation}
where $ \tau _A = \frac{1}{d_{code}} \tr_{\Bar{A}}P_{code} $.
It is good to note here that  the condition
\begin{equation}
     \langle \phi _a \rangle _{\rho_{bulk}} = \langle \Phi _{a,HKLL} \rangle _{\rho_{CFT}}
\end{equation}
implies that 
$ V_{HKLL}(\phi _a) V_{HKLL}^\dagger = P_{code}  \Phi _{a,HKLL} P_{code} $, and so the bulk operator $ \phi$ in the entanglement wedge can map to a boundary operator has support only in the region $A$ as
\begin{equation}
     O_A
    =\frac{1}{d_{code}}
    \tau _A ^{-1/2}
     \tr_{\Bar{A}} \big(P_{code}  \Phi _{HKLL} P_{code}\big)
  \tau _A^{-1/2}.
\end{equation}
This is the main result of the section, that we will use in the rest of the paper.

When including $1/N$ corrections, \eqref{13} will no longer be exact and we do not expect to have an exact recoverability. In that case, we can try to reconstruct the entanglement wedge using the twirled Petz map \eqref{4}. For the maximally mixed state in the code subspace, the mapping is as below
\begin{equation}
     O_A= \frac{1}{d_{code}} 
    \int _\mathbb{R} dt ~\beta_0 (t) ~
    \tau_A ^{-1/2(1-it)}
    \tr_{\Bar{A}} \big(V_{HKLL}(\phi _a) V_{HKLL}^\dagger\big)
    \tau_A ^{-1/2(1+it)}
\end{equation}
which at large $N$ limit gives us the same formula as \eqref{18} \cite{cotler2019entanglement}. It has been argued that for the reconstruction of the entanglement wedge for any finite-dimensional code subspace as well as code subspaces with dimensions that do not grow exponentially fast in $N$, while the error is non perturbatively small, the ordinary Petz map works well enough \cite{chen2020entanglement}.

In the large $N$ limit it is possible to take the size of the code subspace to infinity. In that case, the maximally mixed state on the code subspace does not really exist and we would need to use some regulated version of it\footnote{For example, this could be a thermal state, which approximates the maximally mixed state as $T\rightarrow\infty$.} that we denote by $\rho$. 

One should be careful at this point that the quantum channel in 
(\ref{16}), which takes as input the reduced density matrix of the entanglement wedge $\rho _a = \tr_{\Bar{a}}(\rho)$ and gives as output a state on $A$,  will not  generally provide us exactly the same state on $A$ as  
$ \rho _A = \tr_{\Bar{A}} \big( V_{HKLL}(\rho) V_{HKLL}^\dagger\big) $ which depends on the state $ \rho$ defined on the entire bulk. 
Only in the case that the bulk reference state itself is a tensor factor of two states in $a$ and $\Bar{a}$, like the maximally mixed state, they will give us the same result. 
However, their difference is controlled by $ 1/N$: if we say that
   $ | S(\rho_ A|| \sigma _A )-S(\rho_ a|| \sigma _ a)|\leqslant \epsilon,
$
then
\begin{equation}
   \| \mathcal{E}(\rho _a) - \rho _A\|^2 \leqslant 2 \ln 2 S(\mathcal{E}(\rho _a) || \rho _A)\leqslant ( 2 \ln 2) \epsilon .
\end{equation}
Hence, at large $N$ limit that $\epsilon$ goes to zero and we have the exact reconstruction of the entanglement wedge, one can exchange the $ \mathcal{E}(\rho _a) $ and $  \rho _A$. 
Then, we can introduce a general version of the Petz map in terms of an arbitrary fixed state $\rho$ as \cite{penington2019replica}
\begin{equation}
     O^{(\rho)}_A
    =
    \rho _A ^{-1/2}
     \tr_{\Bar{A}} \big( V_{HKLL}(\rho ^{1/2}\phi _a \rho ^{1/2}) V_{HKLL}^\dagger \big)
  \rho _A^{-1/2}.
\end{equation}
We note here that, for this reconstruction, the only source of the error is not the $1/N$ correction, but rather the entanglement in the state $\rho$ between the inside and outside of the entanglement wedge causes to not recover the original state.

\section{AdS-Rindler reconstruction and Petz map}

As we saw in the previous chapters, a free scalar field in pure AdS is dual to a GFF of the boundary that can be thought of as a sector of a much larger CFT with a large central charge.
In addition, 
Petz map is a tool that comes from the quantum information theory which provides us the CFT representation of the bulk field $ \phi(X)$ that is localized in any region $A$ when the field lies in the entanglement wedge of $A$. It is given by
\begin{equation}\label{42}
    \Phi_A(X)= \tau_A^{-1/2} \tr_{\Bar{A}} \big( P_{code}  \Phi_{HKLL}(X)  P_{code}\big) \tau_A^{-1/2} 
\end{equation}
where we redefine $ \tau_A$ to the unnormalized maximally mixed state  $ \tau_A =\tr_{\Bar{A}}  P_{code} $ and $\Phi_{HKLL}(X)$ is the boundary reconstruction of the field in global coordinates
\begin{equation}\label{43}
    \Phi_{HKLL}(X) = \int _{bdy} dt d\Omega ~ K^g(X|t,\Omega) O(t,\Omega)
\end{equation}
that $ K^g(X|t,\Omega)$ is the smearing function for the AdS space which in even and odd dimension given  by (\ref{40}) and (\ref{41}) respectively. 
By plugging (\ref{43}) back into (\ref{42}) and considering the linearity of the trace we will get
\begin{equation}\label{48}
      \Phi_A(X)= \int _{bdy} dt d\Omega~  K^g(X|t,\Omega)~
      \tau_A^{-1/2} \tr_{\Bar{A}} \big( P_{code}  O(t,\Omega)  P_{code}\big) \tau_A^{-1/2}.
\end{equation}
Therefore, to find $ \Phi_A(X) $ we need to deal with terms
\begin{equation}\label{44}
     \tr_{\Bar{A}} \big( P_{code}  O(t,\Omega)  P_{code}\big)
\end{equation}
for every  $ O(t,\Omega)$. In order to take trace over $\Bar{A}$, we need to re-express them in terms of the operators that act just on the Cauchy surface $ \Sigma$. In other words, we should use the Heisenberg picture and rewrite all $ O(t,\Omega)$ in terms of the scalar primaries on $\Sigma$ by evolving them with boundary Hamiltonian. Let us consider $ \Sigma$ to be $t=0$ slice. Then, (\ref{44}) can be read off as
\begin{equation}\label{45}
     \tr_{\Bar{A}} \big( P_{code} e^{i H_{CFT}t} O(\Omega) e^{-i H_{CFT}t}  P_{code}\big) =  \tr_{\Bar{A}} \big( P_{code} e^{i H_{GFF}t} O(\Omega) e^{-i H_{GFF}t}  P_{code}\big).
\end{equation}
Since we project the Heisenberg picture operators on the code subspace, which should be a subspace of the GFF sector of the CFT, the CFT Hamiltonian can be replaced by the Hamiltonian of generalized free theory, which is 
\begin{equation}
   H_{GFF}= \sum _{nlm} \omega_{nlm} O^\dagger _{nlm} O_{nlm}.
\end{equation}
It is important to note that all the operators in (\ref{45}) have support on entire $ \Sigma$, even when $ t=0$ and $ O(x_ A)$ is localized in region $A$, $ P_{code}O(x_A)P_{code} $ still can have support on $ \Bar{A}$. 

To do the calculation, it can be more convenient to go to Fourier space. By substituting (\ref{46}) into (\ref{45})
and plugging it back to (\ref{48}) we arrive to 
\begin{multline}\label{80}
     \Phi_A(X) = \sum _{nlm} G_{nlm}(X) \tau_A^{-1/2}  \tr_{\Bar{A}} \big( P_{code}  O_{nlm} P_{code} \big) \tau_A^{-1/2} 
     \\
      + G^*_{nlm}(X) \tau_A^{-1/2} \tr_{\Bar{A}} \big( P_{code}  O^\dagger_{nlm} P_{code}\big) \tau_A^{-1/2}
\end{multline}
where 
\begin{equation}
    G_{nlm}(X) = \int _{bdy}  dt d\Omega~ K^g (X|t,\Omega)~ g_{nlm}(t,\Omega).
\end{equation}

To go ahead, we need to determine more precisely the setup we want to study and in particular, specify the region $A$ on the boundary.
Let us start with a simple case. Take $ A$ to be just one hemisphere of $\Sigma$, then as a result, $ a= \mathcal{E}_A$ is an AdS-Rindler wedge in the bulk which its entanglement wedge coincides with its causal wedge. In the rest of this chapter, we will focus on finding the boundary representation for the operators that lie in the AdS-Rindler wedge. First, we work on rewriting  the operator $ O_{nlm}$ in terms of the operators that act just on $A$ or $ \Bar{A}$. We will then define the code subspace in this case and in particular, we will try to find a suitable choice of basis for code subspace to be able to do the calculation. Finally, we will compare our result with the boundary representation of the field one can find from the HKLL procedure in the AdS-Rindler coordinates.

\subsection{Boundary treatment}
The computation of \eqref{80} involves tracing out the sub-region $\bar{A}$ from the operator $ P_{code}  O_{nlm} P_{code}$. For this, one needs to choose appropriate basis to express $P_{code}$, discussed in Sec.\ref{codesub}, and rewrite $O_{nlm}$ in a way that makes the tracing out of complement easier. 

Two ways to proceed further are discussed in this and the next section. One way is to use the bulk Bogoliubov transformation between the global and Rindler modes to adopt the same transformation for the global and Rindler boundary modes. Note that the boundary modes, by themselves, do not have Bogoliubov transformation, since they do not satisfy any equation of motion. Thus, to claim such transformation for the boundary modes, one needs to use the AdS-Rinlder reconstruction to relate the bulk Rindler modes to the boundary Rindler modes. 

The other way, which will be discussed in this section, is solely a boundary treatment. As will be seen in Sec. \ref{codesub}, while computing the trace, one is interested in computing matrix elements like $\bra{\Psi}O_{nlm}\ket{\Psi}$ where $\ket{\Psi}$ is an a state in the code subspace, 
where $O_{nlm}$ is a global mode of single trace operators localized in full boundary. One can choose a Cauchy surface, say at $t=t_{0}$, that includes the bulk operator and after dividing its boundary into two subregions, $A$ or $\bar{A}$, and write single trace operators localized  in the two subregions which we call $O(\tau=0,r)_{A}$ and $O(\tau=0,r)_{\bar{A}}$ respectively (in general one has single trace operators in the wedges associated to the two subregions and in the past and future wedges). Given that the subregion $A$ is spherical, one can write the Fourier transform of $O(\tau,r)_{A}$, the single trace operator in the domain of dependence of $A$, as    
\begin{equation}\label{O}
  O(\tau, x)_{A}  = \int d\omega d\lambda e^{-i\omega\tau} Y_\lambda (x) O_{\omega\lambda, A} + h.c.
\end{equation}
where $Y_\lambda (x)$ is the eignefunction of the Laplacian on a co-dimension one hyperbolic ball in the boundary.


In addition, due to the particle number conservation of AdS/CFT in the large N limit, the action of the normalized operator $O_{nlm}/M_{nlm}$\footnote{The normalization can be computed from the two point function of $O^{\dagger}_{nlm}$ on the vacuum, requiring that $\frac{O^{\dagger}_{nlm}}{M_{nlm}}\ket{\Omega}$ has norm one.}  on any state in the code subspace is given by a linear combination of the single trace operators in domain of dependence of $A$ and $\bar{A}$, on the state after being appropriately normalized. One may expect that similar Rindler modes in the past and future wedges should also contribute but 
it can be checked, from 
the boundary, that the modes in the future and past wedges can be expressed in terms of modes in the domain of dependence of $A$ and $\bar{A}$ by the symmetry property of the problem and requiring the correct two point functions in the future and past wedges.
Thus one can just replace $P_{code}O_{nlm}P_{code}$ inside the equation by 
\begin{equation}\label{l}
\sum_{I=A,\bar{A}} M_{nlm} \sum_{\omega \lambda} P_{code} \frac{1}{M_{\omega \lambda}} (\chi^{1}_{nlm,\omega \lambda}O_{\omega \lambda,I}+\chi^{2}_{nlm,\omega \lambda}O^{\dagger}_{\omega \lambda,I})P_{code} 
    \end{equation}
  where $M_{\omega \lambda}$ is the normalization such that $O_{\omega \lambda}/M_{\omega \lambda}\ket{\Omega}$ has norm one. The coefficients $\chi^{1,2}_{nlm,\omega \lambda}$ at least have to satisfy
  \begin{equation}
  \sum _{\omega \lambda} (  \chi^1_{nlm; \omega\lambda}\chi^{*1}_{nlm; \omega\lambda}+\chi^1_{nlm; \omega\lambda}\chi^{*2}_{nlm; \omega\lambda}+
  \chi^2_{nlm; \omega\lambda}\chi^{*1}_{nlm; \omega\lambda}+  \chi^{*2}_{nlm; \omega\lambda} \chi^{2}_{nlm; \omega\lambda}) = 1.
\end{equation}
In the next section, one can see that the coefficients $\chi^{1,2}_{nlm,\omega \lambda}$ are indeed the bulk global to Rindler Bogolibov coefficients.

\subsection{Bogoliubov coefficients from Rindler mode expansion of bulk field}

Now we proceed with the second way of arriving at \eqref{l}. At every Cauchy surface, the Hilbert space of a QFT is constructed as the Fock space obtained from creation and annihilation operators $a_k$ and $a^\dagger_k$, corresponding to the global modes of the field operator which is
\begin{equation}\label{61}
    \phi(t, x) = \sum _k f_k(t,x) a_k + f^*_k(t,x) a^\dagger_k.
\end{equation}
$k$ is a collection of indices we need to describe the mode. 
We can use the same approach to find the mode expansion of the field that lies in the region $r$ by directly solving the equation of motion just in this region to find the appropriate wave functions which have support only on $r$. 
Let us take the time slice $\Sigma$ and decompose it into the subregions $\Sigma_r$ such that $ \Sigma_r \cap \Sigma_{r'} = \varnothing $. For all $ \Sigma_r$, we should first find  a coordinate system $U_r$ which cover $ \mathcal{D}(\Sigma_r)$. Then, solve the equation of motion on $ U_r$ to find the mode expansion of fields on $ \mathcal{D}(\Sigma_r)$
\begin{equation}
    \phi(t_r,x_r) =\sum _k f^r_k(t_r,x_r) a^r_k + f^{r*}_k(t_r,x_r) a^{r \dagger}_k.
\end{equation}
The Hilbert space of the QFT restricted to $\Sigma_r$ is denoted by $ \mathcal{H}_r$ and the Hilbert space of the total theory on $\Sigma$ is naively a tensor product of the subregion Hilbert spaces $ \mathcal{H} = \otimes_r \mathcal{H}_r$.

One can expand the field $\phi (X)$ in global coordinates in terms of subregion mode functions as
\begin{equation}\label{62}
    \phi(t,x) = \sum_r\sum _k f^r_k(t_r,x_r) ~a^r_k + f^{r*}_k(t_r,x_r) ~a^{r \dagger}_k.
\end{equation}
The point $X$ is labeled in global coordinates and the coordinate system $U_r$ by $(t, x)$ and $(t_r, x_r)$,  respectively.
As a result, the creation and annihilation operators of the full Hilbert space can be written as a linear combination of subregions mode functions and vice versa, by comparing (\ref{61}) and (\ref{62}) which is a generalized version of Bogoliubov transformation \cite{kim2017explicit}. 

Let us come back to our problem. To proceed in the Petz map calculation, it can help us if we could find an expression for $ O_{nlm}$ in terms of the mode function corresponding to the subregions $A$ and $\Bar{A}$. The subtlety here is the point that GFF on the boundary do not obey the equation of motion and hence, the discussion above is not applied to the boundary QFT. 
However, in AdS/CFT correspondence,  the extrapolate dictionary leads us to the identification between some bulk and boundary operators. As a result, we expect that bulk Bogoliubov transformation can help us to find one  expression  for $O_{nlm}$ as a linear combination of the operators has support only on one subregion.

The boundary Cauchy slice $ \Sigma$ is divided into two hemispheres $ A$ and $ \Bar{A}$. As a result, their entanglement wedges are AdS-Rindler patches in the bulk which both together cover the entire AdS space. To quantize the free fields in the entire AdS space in Rindler coordinates, we need two copies of the creation and annihilation operators that obey the commutation relation 
\begin{equation}\label{51}
    [b _{\omega\lambda, I}, b^{\dagger } _{\omega '\lambda' , I'}] = (2\pi) ^2 \delta (\omega - \omega') \delta (\lambda - \lambda')\delta _{II'}
\end{equation}
where the mode functions $ b _{\omega\lambda, a}$ and $ b _{\omega\lambda, \Bar{a}}$ have support only in $a$ and $ \Bar{a}$ patches respectively. 

One can globally expand the bulk field $ \phi(X)$ in terms of these mode functions as
\begin{equation}
    \phi(X)= \sum _{I \in \{a, \Bar{a} \}}\int \frac{d \omega}{2\pi} \frac{d\lambda}{2\pi} ~ \Big( f_{\omega\lambda,I}(X) b_{\omega\lambda, I}+f^{*}_{\omega\lambda,I}(X) b^\dagger_{\omega\lambda,I}\Big)
\end{equation}
where $f_{\omega\lambda,I}(X) $ is given by (\ref{50}) if the point $ X$ belongs to the patch $I$, otherwise it vanishes.  
The global mode $a_{nlm}$ in AdS are related to these mode functions by Bogoliubov coefficients $ \alpha$ and $\beta$ as
\begin{equation}\label{57}
    a_{nlm} = \sum _{I \in \{a, \Bar{a} \}}\int d\omega d\lambda ~ \Big( \alpha^I _{nlm; \omega\lambda}  b_{\omega\lambda, I} + \beta ^{*I} _{nlm ; \omega\lambda}  b^\dagger_{\omega\lambda, I}\Big).
\end{equation}
The commutation relations (\ref{51}) lead to the following constrain on the Bogoliubov coefficients
\begin{equation}
   \sum _{I \in \{a, \Bar{a} \}}\int d\omega d\lambda ~\Big( \alpha^I_{nlm; \omega\lambda}\alpha^{*I}_{n'l'm'; \omega'\lambda'} -  \beta^{*I}_{nlm; \omega\lambda} \beta^{I}_{n'l'm'; \omega'\lambda'}\Big) = \delta _{nn'} \delta _{ll'}\delta _{mm'}.
\end{equation}
We can substitute (\ref{57}) in the bulk global mode expansion (\ref{60}) which lead us to the relations
\begin{equation}\label{83}
    \begin{split}
      \sum _{nlm} f_{nlm}(X) \alpha^a_{nlm; \omega\lambda} + f^*_{nlm}(X) \beta^a_{nlm;\omega\lambda} =0 \qquad \forall X \in \Bar{a}\\
      \sum _{nlm} f_{nlm}(X) \alpha^{\Bar{a}}_{nlm;\omega\lambda}+ f^*_{nlm}(X) \beta^{\Bar{a}}_{nlm;\omega\lambda} =0 \qquad \forall X \in a.
    \end{split}
\end{equation}
We will use them in what follows.

For the case that we are studying, where on the boundary of pure AdS  the GFF lives, the mode functions $ a_{nlm}$ and $ b_{\omega\lambda}$ are identified with the boundary operators given by (\ref{55}) and (\ref{56}) respectively. 
By plugging them back into (\ref{57}), one can find
\begin{equation}\label{65}
     O_{nlm} = \sum _{I \in \{A, \Bar{A} \}}\int d\omega d\lambda \Big( \frac{M_{nlm}}{M_{\omega\lambda}}\alpha^I _{nlm;\omega\lambda} O_{\omega\lambda, I} +  \frac{M_{nlm}}{M_{\omega\lambda}}\beta ^{*I} _{nlm;\omega\lambda}  O^\dagger_{\omega\lambda, I}\Big).
\end{equation}
while $ \alpha^A _{nlm;\omega\lambda}=\alpha^a _{nlm;\omega\lambda}$, etc .

\subsection{Appropriate basis for the code subspace}\label{codesub}

The code subspace has a Fock space structure
$ \mathcal{H}_{code} = span \{ \prod _{nlm} ( O_{nlm}^{\dagger })^{i_{nlm}}\ket{\Omega}  \}$, where $ \ket{\Omega}$ is the global vacuum defined as $ O_{nlm} \ket{\Omega}=0$ for all $n,l$ and $m$.
The powers $ i_{nlm}$ are some non-negative integers and we can also put a cut-off on them.
In order to compute the Petz map reconstruction of the bulk field $ \phi (X)$ that lies in the AdS-Rindler patch, we need to compute the terms $ \tr_{\Bar{A}} P_{code}$  and $ \tr_{\Bar{A}}(P_{code} O_{nlm} P_{code})$. 

Before  going through the calculation, we need to choose a basis for code subspace.
The natural choice one can take is
\begin{equation}
    \ket{\{i_{nlm}\}} \propto \prod \limits _{nlm} ( O_{nlm}^{\dagger })^{i_{nlm}}\ket{\Omega}. 
\end{equation}
In this basis, we should calculate the terms of the form
\begin{equation}\label{66}
    \tr_{\Bar{A}} \Big( ( O_{nlm}^{\dagger })^{i}\ket{\Omega} \bra{\Omega} ( O_{n'l'm'})^{i'} \Big)
\end{equation}
for every arbitrary integers $i$ and $i'$. 
One way to deal with trace can be using (\ref{65}). 
As we know, the action of Rindler modes on $\ket{\Omega}$ in two wedges are related to each other by
\begin{equation}
 \begin{split}
     O_{\omega,\lambda;\Bar{A}} \ket{\Omega} &=e^{\pi \omega}  O^\dagger_{\omega,-\lambda;A} \ket{\Omega} \\
      O^\dagger_{\omega,\lambda;\Bar{A}} \ket{\Omega} &=e^{-\pi \omega}  O_{\omega,-\lambda;A} \ket{\Omega} 
 \end{split}   
\end{equation}
while $  \ket{\Omega} = \otimes_{\omega,\lambda} \ket{\Omega_{\omega,\lambda}}= \otimes _{\omega,\lambda} \sqrt{1- e^{-2\pi \omega}} \sum _n e^{-\pi \omega n} \ket{n}_{\omega,\lambda}^A \ket{\Bar{n}}_{\omega,-\lambda}^{\Bar{A}}$.
As a result, for each choice of $i$,  one can in principle find an operator $ A_i$ that has support only on region $A$ such that 
\begin{equation}\label{68}
   ( O_{nlm}^{\dagger })^{i}\ket{\Omega} = A_i \ket{\Omega} .  
\end{equation}
Therefore, (\ref{66}) can be simplified as
\begin{equation}
       A_i \tr_{\Bar{A}} \big(\ket{\Omega} \bra{\Omega}\big)  A_i^\dagger =   A_i ~\rho ^{(0)}_{A}~A_i^{\dagger}
\end{equation}
that is an operator has support only on $A$, while $ \rho ^{(0)}_{A}$ is a thermal density matrix in the region $A$. Nevertheless, the equation (\ref{68}) is somewhat abstract, and indeed finding an expression for $A_i$ can be difficult.  

To find a more convenient basis for the code subspace we can use the \emph{Reeh-Schlieder} theorem for relativistic QFT. Consider a QFT in Minkowski spacetime $\mathcal{M}$ with a Hilbert space
$ \mathcal{H}$ and the vacuum state denoted by $ \ket{\Omega} \in \mathcal{H} $. For a small open set $ \mathcal{U} \subset \mathcal{M}$, there is a bounded algebra of local operators $ \mathcal{A}_\mathcal{U}$ supported in $ \mathcal{U}$. The Reeh-Schlider theorem says that every arbitrary state in $ \mathcal{H}$ can be approximated by $\mathcal{A}_\mathcal{U} \ket{\Omega} $ that means states created by applying elements of the local algebra to the vacuum are not localized to the region $ \mathcal{U}$. In other words, the vacuum is a cyclic and separating vector for the field algebra corresponding to any open set $\mathcal{U}$ in Minkowski spacetime. 
This is the key point in our work that causes the manageability of the Petz map calculation. 

We can construct the code subspace using the Reeh-Schlieder theorem to the boundary QFT by acting on the global vacuum with the operator algebra on region $A$, $ \mathcal{H}_{GFF} = \{\mathcal{L}(\mathcal{H}_A) \ket{\Omega}\}$. Since one choice of basis for the operator algebra on $A$ is the set of Rindler modes $ O_{\omega\lambda;A}$ and $ O^{\dagger}_{\omega\lambda;A}$, one can take a basis for code subspace at large $N$ as
\begin{equation}
    \ket{\{j_{\omega,\lambda}, \Delta_{\omega,\lambda}\}} = \prod\limits_{\omega,\lambda} 
    (O_{\omega\lambda;A})^{j_{\omega,\lambda}} (O^{\dagger}_{\omega\lambda;A})^{j_{\omega,\lambda}+ \Delta _{\omega,\lambda}} \ket{\Omega} 
\end{equation}
where $ j \in \mathbb{N}$ and $ \Delta \in \mathbb{Z}$. 
\footnote{ More precise statement here is that, since the representation of the vacuum state in terms of the Rindle modes is cyclic and separating with respect to the operator algebra of the Rindler wedge, the vacuum sector of the Hilbert space is isomorphic to the GNS Hilbert space of the operator algebra of the Rindler wedge over the vacuum.}
As the theory is free, different modes are decoupled and we can rewrite the code subspace basis as
\begin{equation}
     \ket{\{j_{\omega,\lambda}, \Delta_{\omega,\lambda}\}} = \otimes _{\omega,\lambda} \ket{j_{\omega,\lambda}, \Delta_{\omega,\lambda}}
     =\otimes_{\omega,\lambda} 
    (O_{\omega\lambda;A})^{j_{\omega,\lambda}} (O^{\dagger}_{\omega\lambda;A})^{j_{\omega,\lambda}+ \Delta _{\omega,\lambda}} \ket{\Omega_{\omega,\lambda}} 
\end{equation}
In the following, for simplicity we will just focus on a single mode which the corresponding Hilbert space is $ span \{ \ket{j, \Delta} = (O_{A})^{j} (O^{\dagger}_{A})^{j+ \Delta} \ket{\Omega}  \} $.
In the new basis, instead of (\ref{66}), we should calculate the terms 
$ \tr_{\Bar{A}}\ket{j, \Delta} \bra{j', \Delta'}$ which one can simply find as 
\begin{equation}
    \tr_{\Bar{A}}\ket{j, \Delta} \bra{j', \Delta'}=(O_{A})^{j} (O^{\dagger}_{A})^{j+ \Delta} \rho ^{(0)}_{A} (O_{A})^{j'+\Delta'} (O^{\dagger}_{A})^{j'}
\end{equation}

We should be careful here that although this set of vectors spans the GFF sector of the boundary,  they are not orthonormal as we have
\begin{equation}
    \langle j, \Delta\ket{ j', \Delta'} =  \delta _{\Delta,\Delta'} (1- e^{-2\pi\omega}) \sum _{n= max\{0, -\Delta\}} e^{-2\omega n} \sqrt{\frac{(n+j+\Delta)!}{(n+\Delta)!}}  \sqrt{\frac{(n+j'+\Delta')!}{(n+\Delta')!}} 
\end{equation}
which is proportional to $\delta _{\Delta,\Delta'}$ not $ \delta _{j,j'}\delta _{\Delta,\Delta'}$. 
Nevertheless, we can still use this set of vectors as a basis for the code subspace by considering the correct  form of the projection on a non-orthonormal basis.

Consider a vector space $ V = span \{ \ket{v_i}\}$. One can construct the metric tensor for this basis  $ G = [g_{ij}]$ that by definition 
$ g_{ij} = \langle v_i \ket{v_j}$. The inverse metric
$ G^{-1} = [g^{ij}]$ 
is defined to be the inverse of the matrix $G$, so the relations 
\begin{equation}\label{82}
        \sum _j g^{ij} g_{jk} = \delta ^i _k,~~~~~~~~~~~~~~~~
        \sum_j g_{ij} g^{jk} = \delta ^k _i
\end{equation}
should satisfy and the projection on the subspace $ V_I = span \{ \ket{v_i}, i\in I\}$ is given by 
\begin{equation}
    P_I = \sum _{i,j \in I} g^{ij} \ket{v_i} \bra{v_j}.
\end{equation}

\subsection{AdS-Rindler wedge reconstruction via Petz map}

Now we have all relations we need to find the Petz reconstruction for the fields in the AdS-Rindler patch in the bulk. 
By plugging (\ref{65}) back into (\ref{80}), we arrive to 
\begin{multline}
    \Phi_A(X) =  \sum _{I \in \{A, \Bar{A} \}}\int d\omega d\lambda ~
    \mathcal{F}_{\omega,\lambda;I}(X)~ 
    \tau_A^{-1/2}  
    \tr_{\Bar{A}} ( P_{code}   O_{\omega\lambda, I} P_{code}) \tau_A^{-1/2} 
    \\
   + \mathcal{F}^*_{\omega,\lambda;I} (X) ~ \tau_A^{-1/2}  \tr_{\Bar{A}} ( P_{code}   O^\dagger_{\omega\lambda, I} P_{code}) \tau_A^{-1/2} 
\end{multline}
while 
\begin{equation}\label{81}
    \mathcal{F}_{\omega,\lambda;I} (X)= \sum _{nlm} \frac{M_{nlm}}{M_{\omega\lambda}} (G_{nlm}(X) \alpha^I _{nlm;\omega\lambda}  +  G^*_{nlm}(X) \beta ^{I} _{nlm;\omega\lambda} ). 
\end{equation}
By comparing the global mode expansion of $ \Phi_{HKLL}(X)$ with $\phi(X)$, one can find that $ G_{nlm} (X)= \frac{1}{M_{nlm}}f_{nlm}(X)$. If we substitute  it in (\ref{81}), we can find that 
\begin{equation}
   \mathcal{F}_{\omega,\lambda;I}(X) = \frac{1}{M_{\omega\lambda}} \sum _{nlm}  (f_{nlm}(X) \alpha^I _{nlm;\omega\lambda}  +  f^*_{nlm}(X) \beta ^{I} _{nlm;\omega\lambda} )
\end{equation}
As $ \phi(X)$ lies in the AdS-Rindler wedge homologous to the region $A$,  by using the relations (\ref{83}), we find that 
$  \mathcal{F}_{\omega,\lambda;\Bar{A}}(X)=0 $ for all $ X \in \mathcal{E}_A$. Therefore, the Petz  reconstruction of $ \phi(X)$ gets simplified as
\begin{multline}
    \Phi_A(X) = \int d\omega d\lambda ~
    \mathcal{F}_{\omega,\lambda;A}(X)~ 
    \tau_A^{-1/2}  \tr_{\Bar{A}} ( P_{code}   O_{\omega\lambda, A} P_{code}) \tau_A^{-1/2}  
    \\
  + \mathcal{F}^*_{\omega,\lambda;A} (X) ~ \tau_A^{-1/2}  \tr_{\Bar{A}} ( P_{code}   O^\dagger_{\omega\lambda, A} P_{code}) \tau_A^{-1/2} .
\end{multline}

In our basis the projection to the code subspace is
\begin{equation}
    P_{code} = \sum _{j,j'} \sum _{\Delta, \Delta'} g^{(j,\Delta);(j',\Delta')} \ket{j, \Delta} \bra{j', \Delta'}. 
\end{equation}
From the inner product between $\{ \ket{j, \Delta}\}$, we see that the metric tensor here is block-diagonal while each block labeled by $\Delta$
\begin{equation}
    G =\oplus _\Delta G_\Delta = \oplus _\Delta [g_{j,j';\Delta}]
\end{equation}
where $ g_{j,j';\Delta} = \langle j, \Delta\ket{ j', \Delta} $.
As a result, the inverse metric should have the form of 
\begin{equation}
    G^{-1} =\oplus _\Delta G^{-1}_\Delta = \oplus _\Delta [A_{j,j'}^{\Delta}]
\end{equation}
for some unknown elements $ A_{j,j'}^{\Delta}$ which 
should satisfy the relations below
\begin{equation}\label{84}
\begin{split}
  \sum _{j'}  A_{j,j'}^{\Delta} \langle j', \Delta\ket{ j'', \Delta}  = \delta _{j,j''}
  \\
  \sum _{j'}  \langle j, \Delta\ket{ j', \Delta}  A_{j',j''}^{\Delta}  = \delta _{j,j''}.
\end{split}
\end{equation}
Since $ g^{(j,\Delta);(j',\Delta')} = A_{j,j'}^{\Delta} \delta_{\Delta, \Delta'}$,
we can write the projection on the code subspace in terms of $ A_{j,j'}^{\Delta} $ as
\begin{equation}
      P_{code} = \sum _{\Delta} \sum _{j,j'} A_{j,j'}^{\Delta} \ket{j, \Delta} \bra{j', \Delta}. 
\end{equation}

Now, we can use the form of the code subspace projection to find the three terms we need to find the Petz reconstruction of $\phi (X)$.
First, we  start with $ \tau _A$ which is
\begin{multline}
        \tau_A = \tr_{\Bar{A}} P_{code} =
    \sum _{\Delta} \sum _{j,j'} A_{j,j'}^{\Delta} \tr_{\Bar{A}}\ket{j, \Delta} \bra{j', \Delta}
    \\
    = 
     \sum _{\Delta} \sum _{j,j'} A_{j,j'}^{\Delta} (O_{A})^{j} (O^{\dagger}_{A})^{j+ \Delta} \rho ^{(0)}_{A} (O_{A})^{j'+\Delta'} (O^{\dagger}_{A})^{j'}.
\end{multline}
We also need to calculate the terms in the form of
\begin{equation}
   \tr_{\Bar{A}} \big(P_{code} O P_{code}\big) =  \sum _{\Delta, \Delta'} \sum _{j,j'} \sum _{k,k'} 
   A_{j,k}^{\Delta} A_{k',j'}^{\Delta'} \bra{k, \Delta} O \ket{k', \Delta'}
   \tr_{\Bar{A}}\ket{j, \Delta} \bra{j', \Delta'},
\end{equation}
For $ O$ that is $ O_A$ or $O_A^\dagger$, we get
\begin{equation}\label{85}
    \begin{split}
      \bra{k, \Delta} O_A \ket{k', \Delta'}= &\bra{k, \Delta} k'+1, \Delta'-1 \rangle
      \\
      \bra{k, \Delta} O_A^\dagger \ket{k', \Delta'}=&\bra{k+1, \Delta-1} k', \Delta' \rangle.
    \end{split}
\end{equation}
By using the relations (\ref{84}) and (\ref{85}), one can find that 
\begin{equation}
    \begin{split}
        \tr_{\Bar{A}} \big(P_{code} O_{\omega,\lambda;A} P_{code}\big) =& O_{\omega,\lambda;A} \tau_A
        \\
         \tr_{\Bar{A}} \big(P_{code} O_{\omega,\lambda;A}^\dagger P_{code}\big) = &\tau_A O_{\omega,\lambda;A}^\dagger.
    \end{split}
\end{equation}
The operators $O_{\omega,\lambda;A}$ and $ O_{\omega,\lambda;A}^\dagger$ have support only on region $A$ and commute with every operator $X_{\Bar{A}}$. We can show that here, it is equivalent to say that $ \tau _A ^ {-1}$ commute with the 
$ \tr_{\Bar{A}} (P_{code} O P_{code})$ for $O$ being $  O_{\omega,\lambda;A}$ and $ O_{\omega,\lambda;A}^\dagger$. One can conclude that if they commute with $ \tau _A ^ {-1}$,
they commute with $ \tau_A ^{-1/2}$ as well. Therefore, we reach
\begin{equation}
    \begin{split}
        \tau_A ^{-1/2}\tr_{\Bar{A}} \big(P_{code} O_{\omega,\lambda;A} P_{code}\big) \tau_A ^{-1/2}=&
        \tr_{\Bar{A}} \big(P_{code} O_{\omega,\lambda;A} P_{code}\big) \tau_A ^{-1}=
        O_{\omega,\lambda;A} 
        \\
          \tau_A ^{-1/2}\tr_{\Bar{A}} \big(P_{code} O_{\omega,\lambda;A}^\dagger P_{code}\big) \tau_A ^{-1/2} =&
         \tau_A ^{-1} \tr_{\Bar{A}} \big(P_{code} O_{\omega,\lambda;A}^\dagger P_{code}\big) = O_{\omega,\lambda;A}^\dagger.
    \end{split}
\end{equation}
Finally, we find the Petz reconstruction of the bulk field $ \phi(X)$ in the AdS-Rindler wedge as
\begin{equation}\label{86}
     \Phi_A(X) = \int d\omega d\lambda ~ \Big(
    \mathcal{F}_{\omega,\lambda;A}(X) ~ O_{\omega\lambda, A} + 
   \mathcal{F}^*_{\omega,\lambda;A} (X) ~  O^\dagger_{\omega\lambda, A} \Big).
\end{equation}
By substituting (\ref{34}) in (\ref{86}), we will arrive at
\begin{equation}
    \Phi_A(X) = \int d\tau dx~ K_{Petz, A} (X|\tau, x)~ O(\tau ,x)
\end{equation}
where the smearing function is
\begin{equation}
    \begin{split}
        K_{Petz, A} (X|\tau, x)= &  \int d\omega d\lambda ~
    \mathcal{F}_{\omega,\lambda;A}(X) e^{i\omega\tau} Y^*_\lambda (x)
    \\ =&   \int d\omega d\lambda~ 
     e^{i\omega\tau} Y^*_\lambda (x) ~\frac{1}{M_{\omega,\lambda}}\sum _{nlm}\Big( f_{nlm}(X) \alpha^a_{nlm; \omega\lambda} + f^*_{nlm}(X) \beta^a_{nlm;\omega\lambda}\Big) 
     \\
     = &  \int \frac{d \omega}{2\pi} \frac{d\lambda}{2\pi}~ \frac{1}{M_{\omega,\lambda}}  f_{\omega\lambda,A}(X)  e^{i\omega\tau} Y^*_\lambda (x). 
    \end{split}
\end{equation}
By comparing with (\ref{26}), we see that the result one can find by applying the Petz map 
in an AdS-Rindler patch is exactly the same as the result of the HKLL procedure in the AdS-Rindler coordinate.

\section{Entanglement wedge reconstruction and  Petz map}

In the previous chapter, we used the Petz map to find the CFT reconstruction of a bulk field in the AdS-Rindler wedge. In principle, this approach can be used to reconstruct the entanglement wedge of any region on the boundary explicitly.
Let us consider CFT$_d$ in a semi-classical state $\ket{\Psi}$ which is dual to a smooth asymptotically AdS spacetime $ \mathcal{M}$. We also assume that there is no black hole in the bulk. Consider a Cauchy surface $ \Sigma$ of the boundary and divide it into an arbitrary region $A$ and its complementary part $\Bar{A}$. In the rest, we focus on finding the reconstruction of the entanglement wedge of $A$ via the Petz map.

In the bulk, one can find the global mode expansion of the field $\phi$ as 
\begin{equation}\label{87}
    \phi (X) = \sum _n \Big ( f_n(X) ~a_n + f^*_n (X) ~a^\dagger_n\Big )
\end{equation}
where $f_n(X)$ is the solution of the Klein-Gordon equation on $\mathcal{M}$ and $a_n$ is the mode corresponding to it that obeys the usual canonical commutation relations. All the labels needed to define the modes are shown collectively by $n$.
By applying the HKLL method to an appropriate coordinate system that covers the entire bulk, labeled here by $ (r,t,x)$, one can find that
\begin{equation}
    \Phi_{HKLL}(X) = \int_{bdy} dt dx~ K^g_{\partial\mathcal{M}}(X|t,x) O(t,x)
\end{equation}
where $K^g_{\partial\mathcal{M}}(X|t,x) $ is the global smearing function. As in the AdS-Rindler case, it is convenient to go to the Fourier modes where the single trace primaries have a mode expansion as
\begin{equation}
    O(t,x) = \sum_n  \Big( \tilde{g}_n(t,x)~ \hat{O}_n + \tilde{g}^*_n(t,x)~ \hat{O}^\dagger_n \Big)
\end{equation}
where
\begin{equation}
    \hat{O}_n = \frac{1}{M_{n}}\int dt dx~ O(t,x) ~g_n^*(t,x).
\end{equation}
If we choose $ \hat{O}_n$ with standard commutation relation, i.e. identified it with $a_n$, from extrapolate dictionary, we have 
$ \tilde{g}_n (t,x) = \lim_{r\rightarrow \infty} r^\Delta f_n (r,x)$, where $M_{n}$ and $g_{n}$ are defined as in Sec. \ref{modesum}. Therefore
\begin{equation}\label{88}
     \Phi_{HKLL}(X) =  \sum_n \Big ( G_{n,\mathcal{M}}(X) ~\hat{O}_n + G^*_{n,\mathcal{M}}(X) ~\hat{O}^\dagger_n \Big)
\end{equation}
while
\begin{equation}\label{95}
    G_{n,\mathcal{M}}(X)  = \int_{bdy} dt dx~ K^g_{\partial\mathcal{M}}(X|t,x) \tilde{g}_n(t,x).
\end{equation}
By comparing (\ref{87}) and (\ref{88}), one can find that $ G_{n,\mathcal{M}}(X) = f_n(X) $.

Let us choose a  basis for the operator algebra  of the regions $A$ and $\Bar{A}$ which we denote them  by $ \{A_\nu\}$ and $ \{\Bar{A}_\nu \}$ respectively.
In order to find the Petz reconstruction of $\phi(X)$, we need to write the mode functions $\hat{O}_n$ as a linear combination of  $ \{A_\nu\}$ and $ \{\Bar{A}_\nu \}$. If it is in the form of
\begin{equation}\label{223}
    \hat{O}_n = \sum _\nu \alpha^A_{n,\nu}~ A_\nu + \alpha^{\bar{A}}_{n,\nu} ~\Bar{A}_\nu.
\end{equation}
The Petz reconstruction of $\phi(X)$ arrives to
\begin{equation}\label{224}
   \Phi_A(X) =  \sum _\nu
    \mathcal{F}_{\nu}^A(X) 
    \tau_A^{-1/2}  \tr_{\Bar{A}} \big(P_{code}  A_\nu P_{code}\big) \tau_A^{-1/2} + 
   \mathcal{F}^{\bar{A}*}_{\nu} (X)  \tau_A^{-1/2}  \tr_{\Bar{A}} \big( P_{code}   \bar{A}_\nu P_{code}\big) \tau_A^{-1/2}  
\end{equation}
where
\begin{equation}\label{94}
  \mathcal{F}_{\nu}^I (X) = \sum _n f_n(X) ~ \alpha^I_{n,\nu} + f^*_n(X) ~ \alpha^{I*}_{n,-\nu}
\end{equation}
for $I\in \{A, \bar{A}\}$. 

For a generic choice of basis, the coefficients behind  $ \tau_A^{-1/2}  \tr_{\Bar{A}} ( P_{code}\bar{A}_\nu P_{code}) \tau_A^{-1/2}$ does not vanish like the case of AdS-Rindler in the previous chapter. Hence, we also need to calculate these terms here. Moreover, for a generic case of the basis of the operator algebra, the sets of  $ \{A_\nu\}$ and $ \{\Bar{A}_\nu \}$ do not have in general a simple bulk dual, and therefore, we can not find the  Bogoliubov coefficients in \eqref{223}  from the bulk theory.

For now, let us assume that we can somehow find the Bogoliubov coefficients in \eqref{223}. Then in order to proceed, similar to the AdS-Rindler case, we can use the Reeh-Schlieder theorem and write the code subspace as
\begin{equation}
    \mathcal{H}_{code} = \mathcal{L}(\mathcal{H}_A)\ket{\Psi}.
\end{equation}     
In principle, to find the Petz reconstruction in \eqref{224}, it is needed to 
 know the commutation relation between the operator algebras of the regions $A$,  
and rewrite the action of the operator $ \bar{A}_\nu$ on the state $ \ket{\Psi}$ in terms of the operators in region $A$ on the state, \emph{i.e.}, finding the operator $ O_{A,\nu}$ as a function of $ \{A_\nu\}$ such that 
\begin{equation}
    \bar{A}_\nu \ket{\Psi} =  O_{A,\nu} \ket{\Psi}.
\end{equation}
But practically, whether or not we can explicitly compute all the terms in \eqref{224} depends on the basis we take, and for an appropriate choice of it, we will be able to find an explicit expression for the operator $ \Phi_A(X)$. 

In the following,
we will describe 
an appropriate choice of the sets $\{A_\nu\}$ and $\{\bar{A}_\nu\}$ that by using them, the calculation becomes attainable.
We will see that there is a clever choice of basis, the eigenfunctions of the modular Hamiltonian, that the Petz calculation will get drastically simplified.



\subsection{Petz map and modular flow}

In this section, we will focus on a special choice of basis for operator algebra in the region $A$ and $\bar{A}$ that is the eigenfunctions of modular Hamiltonian of the regions. 

The modular Hamiltonian of a given region $R$ is defined as $ K_R= - \log \rho_R $ where $\rho _R$ is the reduced density matrix of the region $R$. $K_R$ generates an automorphism for the operator algebra $ \mathcal{A}_R$ associated to $ \rho_R$ \cite{haag1992local} as 
\begin{equation}
    A \in \mathcal{A}_R \quad\longrightarrow \quad A_s = e^{iK_R s} A e^{-iK_R s} \in \mathcal{A}_R.
\end{equation}
called \emph{modular flow}. The modular flow originally introduced in the context of the algebraic QFT \cite{takesaki1970tensor, haag1992local, bratteli2012operator, borchers2000revolutionizing, takesaki2003theory} which recently played a key role in using the concepts of quantum information theory in QFT and gravity
\cite{ lashkari2019constraining, lashkari2021modular,sarosi2018modular, blanco2018modular,casini2008relative,
blanco2013localization, faulkner2016modular, balakrishnan2019general, ceyhan2020recovering, lashkari2016modular, lashkari2021modular2, cardy2016entanglement, casini2011towards, blanco2013localization, jafferis2016gravity, koeller2018local,czech2017modular, chen2018modular, belin2018bulk, abt2018properties, faulkner2019modular, czech2019modular, de2020holographic, arias2020modular, erdmenger2020resolving}.

In modular Fourier space, the Fourier transformation of $A_s$ is 
\begin{equation}
    A_\omega = \int _{-\infty}^{\infty} ds~ e^{is\omega} e^{iK_R s} A e^{-iK_R s}.
\end{equation}
The operators $A_\omega$ are the eigenfunctions of modular Hamiltonian $[K_R, A_\omega] = \omega A_\omega$.
They also form a basis for operator algebra on region $R$.
Therefore, we can take the eigenfunctions of the modular Hamiltonian of the both  regions $A$ and $\bar{A}$ as the basis for the corresponding operator algebras on these regions. 

Moreover, as we assume that there is no black hole in the bulk, the entanglement wedge of the complementary part of $A$ in the boundary is the same region as the complementary part of the entanglement wedge of the region $A$ and hence, the union of $a$ and $\bar{a}$ covers the entire Cauchy surface. As a result, we can expand both the bulk and boundary global modes as a linear combination of the modular eigenbasis as
\begin{equation}\label{92}
    \begin{split}
        a_n &= \sum _\omega \alpha^a_{n,\omega} A_\omega^a + \alpha^{\bar{a}}_{n,\omega} A^{\Bar{a}}_\omega
        \\
         \hat{O}_n &= \sum _\omega \alpha^A_{n,\omega} A_\omega^A + \alpha^{\bar{A}}_{n,\omega} A^{\Bar{A}}_\omega.
    \end{split}
\end{equation}
In such a case, we can use the JLMS statement that relates the modular Hamiltonian of a given boundary region $A$ to the modular Hamiltonian of its entanglement wedge $a$ as 
\begin{equation}\label{91}
     K_A = K_a +\frac{Area}{4G} + O(1/N).
\end{equation}
Since the area term in the right hand side of (\ref{91}) is proportional to the identity, both $K_{A(\bar{A})}$ and $K_{a(\bar{a})}$ have the same spectrum and we can identify their eigenfunctions as 
\begin{equation}\label{93}
\begin{split}
     A_\omega^{A}& = A_\omega^a \equiv A_\omega
     \\
      A_\omega^{\bar{A}}& = A_\omega^{\bar{a}} \equiv \bar{A}_\omega.
\end{split}
\end{equation}
Therefore as $ \hat{O}_n = a_n$, by comparing (\ref{92}) and (\ref{93}), we see that both $\hat{O}_n$ and $a_n$ have the same Bogoliubov coefficients
$ \alpha^{A(\bar{A})}_{n,\omega} = \alpha^{a(\bar{a})}_{n,\omega}$.
One can replace it into  (\ref{94}) and find that when we take the eigenfunctions of modular Hamiltonian as the basis of operator algebra of subregions, by definition
\begin{equation}
     \mathcal{F}_{\omega}^{\bar{A}} (X) = \sum _n f_n(X)  \alpha^{\bar{a}}_{n,\omega} + f^*_n(X)  \alpha^{\bar{a}*}_{n,-\omega} =0,\qquad \forall X \in a. 
\end{equation}
Therefore, the Petz reconstruction of $ \phi(X)$ in the entanglement wedge
of the region $A$ can be read off as 
\begin{equation}\label{100}
  \Phi_A(X) =  \sum _\omega
    \mathcal{F}_{\omega}^A(X) ~
    \tau_A^{-1/2}  \tr_{\Bar{A}} \big( P_{code}  A_\omega P_{code}\big) \tau_A^{-1/2}  =  
     \sum _\omega
    \mathcal{F}_{\omega}^A(X) ~A_\omega
\end{equation}
where $ A_\omega$ is the eigenfunction of $K_A$ and $\mathcal{F}_{\omega}^A(X)$ is given by 
\begin{equation}
    \mathcal{F}_{\omega}^A(X) = 
    \sum _n ~ 
    \Big( f_n(X)  \alpha^{a}_{n,\omega} + f^*_n(X)  \alpha^{a*} \Big)
\end{equation}
which by using (\ref{95}), it can be rewritten in terms of the global smearing function  as
\begin{equation}
    \mathcal{F}_{\omega}^A(X) = \int_{bdy} dt dx~ K^g_{\partial\mathcal{M}}(X|t,x) 
    \sum _n \Big(\tilde{g}_n(t,x)  \alpha^{A}_{n,\omega} + \tilde{g}^*_n(t,x)  \alpha^{A*}_{n,-\omega}\Big).
\end{equation}

At this point to write the operator $ \Phi_A$ more precisely, we should know more about the $A_\omega$ themselves.
To leading order in AdS/CFT, the bulk field consists of free fields. For free scalar fields on any region $R$ that  all correlators are fixed by the two-point function, the density matrix is Gaussian and the modular Hamiltonian is bilinear. Its eigenfunctions can be labeled by $\omega$ and $X_S$ where the  coordinates $X_S$ corresponds to a codimension 2 surface $ S\in R$ on one Cauchy slice \cite{faulkner2017bulk}. Therefore, we have 
\begin{equation}
    [K_R , \Phi_\omega(X_S)]=\omega \Phi_\omega(X_S)
\end{equation}
where 
\begin{equation}
    \Phi_\omega(X_S) = \int ~ds~ e^{is\omega} e^{iK_Rs} \phi(X_S) e^{-iK_Rs} \qquad \forall X_S \in S.
\end{equation}

Now, let us consider the free scalar field in the entanglement wedge of the region $A$ and the Cauchy surface as the slice of bulk that intersects with $A$. One clever choice for $S$ can be $A$ itself. By using (\ref{91}) and the identification on the boundary 
$ \phi_0(x_A) = O(x_A)$, we can find the modular eigenfunction of $ K_A$ as
\begin{equation}
     O_\omega(x_A) = \int ~ds~ e^{is\omega} e^{iK_As} O(x_A) e^{-iK_As} \qquad \forall x_A\in A.
\end{equation}
By substituting it in (\ref{100}), we find the Petz reconstruction of $\phi(X)$ lies in the entanglement wedge of $A$ as 
\begin{equation}\label{102}
    \Phi _A(X)= 
    \int_{-\infty}^{\infty} ds \int _A dx_A ~K_{Petz, A} (X|x_A,s)~ e^{iK_As} O(x_A) e^{-iK_As}
\end{equation}
where the smearing function is given by
\begin{equation}\label{101}
     K_{Petz, A} (X|x_A,s)= \sum_n \int d\omega~ e^{is\omega}~ \Big(f_n(X) \alpha_n^A (\omega, x_A)+f^*_n(X) \alpha_n^{A*} (-\omega, x_A )\Big)
\end{equation}
while $ \alpha_n^A (\omega, x_A)$ is the Bogoliubov coefficient between $\hat{O}_n$ and $O_\omega(x_A)$.
As we mentioned in (\ref{sec. Heisenberg picture}), the equation (\ref{102}) has been conjectured in \cite{jafferis2016relative}, and also derived in \cite{faulkner2017bulk} through acting with the modular flow on the extrapolate dictionary. 
Here, we could again obtain it by using the Petz map formula which is a more generic approach.  


As a consistency check, let us calculate (\ref{101}) for the AdS-Rindler wedge. In this patch, the modular parameter is just the Rindler time $\tau$ and the modular Hamiltonian is the Rindler Hamiltonian $H_\tau$. To find the smearing function in (\ref{101}), we need to find the Bogoliubov coefficients between $\hat{O}_{nlm}$ and 
$ O_\omega(x_A) = \int d\tau e^{i\omega\tau} O(\tau , x_A)$ which is
\begin{equation}
    \begin{split}
        \alpha_{nlm}^A (\omega, x_A) = \int d\lambda \frac{1}{M_{\omega,\lambda}} Y^*_\lambda(x_A) \alpha _{nlm;\omega,\lambda} \qquad \forall \omega\geq0
        \\
         \alpha_{nlm}^A (\omega, x_A) = \int d\lambda \frac{1}{M_{\omega,\lambda}} Y^*_\lambda(x_A) \beta^* _{nlm;\omega,\lambda} \qquad \forall \omega<0
    \end{split}
\end{equation}
By plugging it into (\ref{101}), we get
\begin{equation}
    \begin{split}
         K_{Petz, A} &(X|x_A, \tau)
         \\
        & = \sum_{nlm} \int d\omega e^{i\omega\tau}
         \int d\lambda \frac{1}{M_{\omega,\lambda}} Y^*(x_A)
         \Big(f_{nlm}(X) \alpha_{nlm;\omega,\lambda}^A+f^*_{nlm}(X) \beta_{nlm;\omega,\lambda}^A\Big)
         \\ 
         &
          =\int \frac{d \omega}{2\pi} \frac{d\lambda}{2\pi} \frac{1}{M_{\omega,\lambda}}  f_{\omega\lambda,A}(X)  e^{i\omega\tau} Y^*_\lambda (x)
    \end{split}
\end{equation}
which is exactly the smearing function that we know from AdS-Rindler wedge reconstruction.

As illustrated, to reconstruct the operator in the interior of the entanglement wedge, we need to learn more about the modular Hamiltonian of general regions in QFTs.

\section{Discussion}

\begin{figure}[h]
    \centering
    \includegraphics[width=0.9\textwidth]{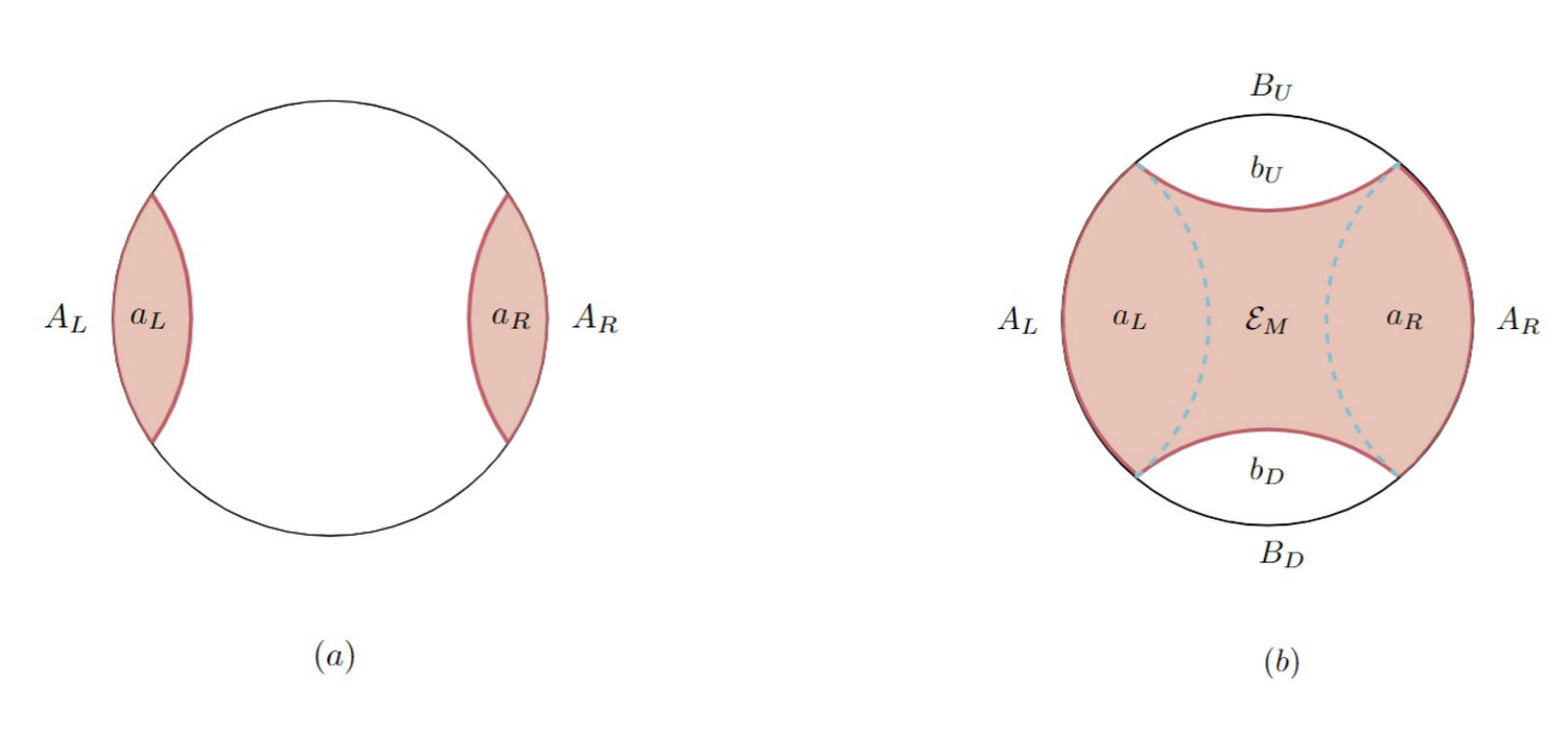}
    \caption{
    The entanglement wedge of a two disjoint intervals $A=A_{L} \cup A_{R}$ in AdS$_3$/ CFT$_2$.
    (a) The entanglement 
    is the region bounded by the boundary region $A$ and the minimal area co-dimension 1 surface
    in the bulk, with the same boundary as $A$. Thus $ \mathcal{E}_A = a_{L}\cup a_{R}$. (b) As one increases the sizes of $A_{L}$ and $A_{R}$, the minimal area surface changes and the entanglement wedge is no longer just $a_{L}\cup a_{R}$, rather it is all of the shade region, that is $\mathcal{E}_A=
    a_{L}\cup a_{R}\cup \mathcal{E}_{M}$. }
    \label{fig:mesh1}
\end{figure}

The discussion of EWR in the last section is generic and applies to any desired region on the boundary. In particular, the region can even be disconnected. For example, let us consider the union of two disjoint intervals
$ A =A_L \cup A_R $, Fig. \ref{fig:mesh1}, on a Cauchy slice of a 2d holographic CFT dual to  AdS$_3$ in the bulk.
If the regions $A_L,~A_R$ are sufficiently small, the entanglement wedge of $A$ is union of the entanglement wedges of $A_L$ and $A_R$, denoted by $ a_L$ and $ a_R$, individually, i.e. the union of two AdS-Rindler wedges. It is well known \cite{almheiri2015bulk,PhysRevD.82.126010} that as we increase the size of the region $A$, the extremal surface changes discontinuously and in the new configuration the entanglement wedge of $A$ becomes larger and in particular larger than the causal wedge of $A$.

An important question is  understanding the nature of observables in the region which is in the entanglement wedge, but not the causal wedge. The Petz formula gives in principle a CFT representation  of these observables, but their microscopic nature is not understood. To make the question more precise, notice that from the point of view of the bulk there is a well defined Bogoliubov transformation between the bulk global modes and 
the modes in regions $a_L,~a_R,~ b_U, ~b_D, ~ \mathcal{E}_{M}$ (see Fig. \ref{fig:mesh1}) . 
The modes in $a_{R},~a_{L},~b_U,~b_D$ can be related to modes of single trace operators in the corresponding boundary regions $A_L,~A_R,~B_U,~B_D$. Entanglement wedge reconstruction and the Petz formula suggests that the modes $d$, which we take to be localized only in $\mathcal{E}_{M}$, should also be representable in region $A=A_L\cup A_R$, but the nature of these observables remains mysterious. Of course the modes $d$ are precisely the modes which are in the entanglement wedge but not the causal wedge of $A=A_L\cup A_R$.

One possibility is that the $d$ modes in region $\mathcal{E}_{M}$ are combinations of complicated operators in region $A_L$ and $A_R$
$$
d= \sum_{ij} c_{ij} ~O^{A_L}_i \otimes O^{A_R}_j
$$
where $O_i^{A_{L,R}}$ are complicated gauge invariant operators. By \emph{complicated} we mean that they are not single-trace or low-multi-trace operators. In this scenario, while each $O_i^{A_{L,R}}$ by themselves do not behave like GFFs, the particular combination above is expected to behave like a GFF in the large $N$ limit.

Another intriguing possibility is that the modes $d$ are operators which are gauge invariant, but they are made out of constituents in regions $A_L,~A_R$ which are not separately gauge invariant. This seems natural from the point of view of, for example,  the free $O(N)$ model. In that case we have operators like $\sum_{L,R} \phi^i(x) \phi^i(y)$, with $x\in A_L$ and $y\in A_R$, which are $O(N)$ invariant but the individual constituents are not, see also discussion in \cite{Mintun:2015qda}.

A difficulty with the second possibility is that in a proper gauge theory
 one would expect that non-gauge invariant operators in regions $A_L,~A_R$ have to be connected by Wilson lines which will have to go through the regions $B_U$ or $B_D$.\footnote{In the $O(N)$ model the symmetry is global hence no Wilson line is necessary.} If the operators in $\mathcal{E}_{M}$ are actually dual to gauge invariant Wilson line operators with end points in $A_L$ and $A_R$, this would imply that they cannot strictly commute with all operators in regions $B_U $ and $ B_D$, as generally the Wilson lines can be detected by operators in regions $B_U$ or $B_D$. This seems to contradict the conventional understanding of EWR, as in the scenario described above the operators on $\mathcal{E}_{M}$ would not be entirely supported in region $A_L,A_R$ since the Wilson lines are passing through the complementary regions.

 
 It would be interesting to explore whether a particular combination of such Wilson lines connecting the individual non-gauge invaritant constituents can be constructed, where commutators of this combination with all simple operators in region $B_U,B_D$ are sufficiently suppressed at large $N$. This might not directly contradict the arguments supporting EWR. For example, the equality of relative entropies \cite{jafferis2016relative} has been established at large $N$ and the arguments are not expected to generalize to imply  equality including exponentially suppressed corrections\footnote{For example, at finite $N$ we expect that the bulk geometry is fully quantum and it is not even clear how one can define the entanglement wedge.}. This might suggest a refinement of EWR where bulk operators are \emph{mostly} supported in $A_R\cup A_L$, allowing some form of Wilson lines connecting the two regions.
 

In any case, the nature of observables in the entanglement wedge but not the causal wedge, like the operators $d$ in this case, remains somewhat mysterious and further study of their properties is necessary.

\acknowledgments

We would like to thank E. Tonni for useful discussions.
In particular, we are greatly indebted to our advisor, K. Papadodimas for the valuable discussions throughout  this work and his useful comments on the preliminary draft as well.
We would also like to acknowledge F. Benini and M. Bertolini for their support as our internal advisors during this work. 
NV would greatly like to thank M. Bertolini and M. Serone for their invaluable support.
The research is partially supported by INFN Iniziativa Specifica - String Theory and Fundamental Interactions project.

\appendix

\section{ HKLL reconstruction in global and Rindler coordinates}\label{A}
 We review here the HKLL reconstruction in global and AdS-Rindler coordinates \cite{hamilton2006holographic} where authors constructed the smearing functions based on the  mode sum approach.
 
\subsection{HKLL reconstruction in global coordinates}

Before going through it, one point that one might be interested in is  if there is any possibility to find a smearing function that has compact support on the boundary of AdS. In particular, we are interested the smearing function has support only on the points that are spacelike separated from $\phi(X)$. The HKLL method provides us with a way of reconstruction in the large $N$ limit where the field $ \phi $ satisfies the free equation of motion.
  Therefore, the smearing function can be constructed from a suitable Green's function that by definition satisfies
\begin{equation}\label{28}
    (\Box - m^2) G(X|X')= \frac{1}{\sqrt{-g}}~ \delta^{d+1}(X-X').
\end{equation}
Using the third Green identity, the field $ \phi$ can be written in global coordinates as
\begin{equation}\label{222}
    \phi (X') = \int dx \sqrt{-g}~ \Big( \phi (X) \partial ^\rho G(X|X') - G(X|X') \partial ^\rho \phi(X)\Big)\Big|_{\rho = \rho_0}
\end{equation}
where $ X= (\rho, x)$, and by sending $ \rho_0\rightarrow \pi /2 $, one can find the smearing function in
(\ref{222}) in terms of the Green's function. For this purpose, let us take the ansatz of Green's function that is non-zero only at spacelike separation \begin{equation}\label{27}
    G(X|X') = f(\sigma (X|X')) \theta (\sigma (X|X')-1),
\end{equation}
where $\sigma$ is an AdS-invariant distant function which in global coordinates is 
\begin{equation}
    \sigma (X|X') = \frac{cos(t-t') - sin(\rho) sin( \rho') cos(\Omega-\Omega')}{cos( \rho) cos (\rho')}
\end{equation}
and $\Omega-\Omega'$ is the angular separation on the sphere. The points that can be connected by a geodesic necessarily lie in the unit cell $ -\pi < t -t' < \pi$. Spacelike separated points are the ones with $ \sigma > 1$ that connected by a geodesic proper distance. By plugging back the ansatz (\ref{27}) to (\ref{28}), we can see that $ f(\sigma)$ satisfies the AdS wave equation. Then, if we start from the beginning  by the ansatz (\ref{27}) \cite{hamilton2006local}, we can find the smearing function with compact support only at spacelike separated region. We note here that this result has been found in global coordinates and in general, it could not be the case. For example, for odd-dimensional AdS  in Poincare coordinates, the smearing function can have support only on the entire boundary. 
 
 Now, let us find  the smearing function in global coordinates. The exact form of the smearing function depends on the dimension. The scalar field solution can be expanded as a linear combination of independent modes that in global coordinates it is given by (Eq. \ref{30}). We can split the field into positive and negative frequency components 
\begin{equation}
    \phi(X) =  \phi(X)_{+} +  \phi(X)_{-}
\end{equation}
where 
\begin{equation}
     \phi(X)_{+} =  \phi(X)_{-} ^\dagger = \sum _{nlm} f_{nlm} a_{nlm}
\end{equation}
and for the boundary operator $ O(x)$ as well. 
Since we can use the AdS isometries to bring one point to another one, it is enough to find the smearing function just at one point. In the center ($\rho =0 $), $ f_{nlm}$ vanishes for all $ l\neq 0$, therefore  only the s-waves contributes to the field in the center of Ads which will simplify the calculation drastically. Let us take $ a_n = a_{n00} = \hat{O}_{n00}$, we can read $a_n$ in terms of $ O(x)$ as
\begin{equation}
    a_n = \frac{1}{\pi vol(S^{d-1}) P_n^{(\Delta -d/2, d/2 -1))} (1)} \int_{-\pi /2}^{\pi /2} dt \int d\Omega \sqrt{g_\Omega}~ e^{i(2n+\Delta)t} O_{+}(t,\Omega).
\end{equation}
Plugging it back into the bulk mode expansion, one can find the bulk field at the origin as
\begin{equation}
    \phi (t', \rho'=0, \Omega') = \int_{-\pi /2}^{\pi /2} dt \int d\Omega \sqrt{g_\Omega}~ K_{+}(t', \rho'=0, \Omega'|t,\Omega) O_{+}(t,\Omega) + h.c.
\end{equation}
where
\begin{equation}
    K_{+}(t', \rho'=0, \Omega'|t,\Omega) =\frac{1}{\pi vol(S^{d-1})} e^{i\Delta t} F(1, d/2, \Delta-d/2+1, e^{-i2t}).
\end{equation} 
It is important to know that smearing functions are not necessarily unique. It can be shifted by terms which vanish when integrated against the boundary operators. It can happen in cases that the boundary fields do not involve a complete set of Foureier modes. This freedom enable us to find $ K_{+}$ which is real and then we can find the kernel such that $ K= K_{+} =K_{-}$. 

Finally, for an arbitrary bulk point by action of an isometry map, we have
\begin{equation}
    \phi (X) = \int_{x\in bdy} dx~ K(X|x) O(x)
\end{equation}
which for the AdS$_{d+1}$ in even dimension, the smearing function is
\begin{equation}\label{40}
    K_G(X|x) = \frac{\Gamma(\Delta-d/2+1)\Gamma(1-d/2)}{\pi vol(S^{d-1})\Gamma(\Delta-d+1)}~ \ lim _{\rho \rightarrow \pi/2} (\sigma(X|x) cos \rho)^{\Delta-d} \theta(\sigma(xX|x)-1)
\end{equation}
and in odd dimension it is given by
\begin{multline}\label{41}
     K_G(X|x) = \frac{2(-1)^{d/2-1}\Gamma(\Delta-d/2+1)}{\pi vol(S^{d-1})\Gamma(\Delta-d+1)\Gamma(d/2)}\\ \lim _{\rho \rightarrow \pi/2} (\sigma(X|x) cos \rho)^{\Delta-d} \log (\sigma(X|x) cos\rho)\theta(\sigma(X|x)-1).
\end{multline}

\subsection{HKLL reconsreuction in AdS-Rindler coordinates}

Consider a CFT Cauchy surface $\Sigma$ and devide it into two regions $A$ and its complementary part $\Bar{A}$. The domain of dependence $ \mathcal{D}(A)$ of $A$ which is the set of points on the boundary that every inextendible causal curve that passes through it must also insert $A$.
The causal wedge of a CFT subregion $A$ is defined as 
\begin{equation}
    \mathcal{C}_A = \mathcal{J}^{+} [\mathcal{D}(A)] \cap \mathcal{J}^{-} [\mathcal{D}(A)]
\end{equation}
where $ \mathcal{J}^{\pm} [R]$ is the bulk causal future/past of region $R$ in the boundary.

\begin{figure}[h]
    \centering
    \includegraphics[width=1\textwidth]{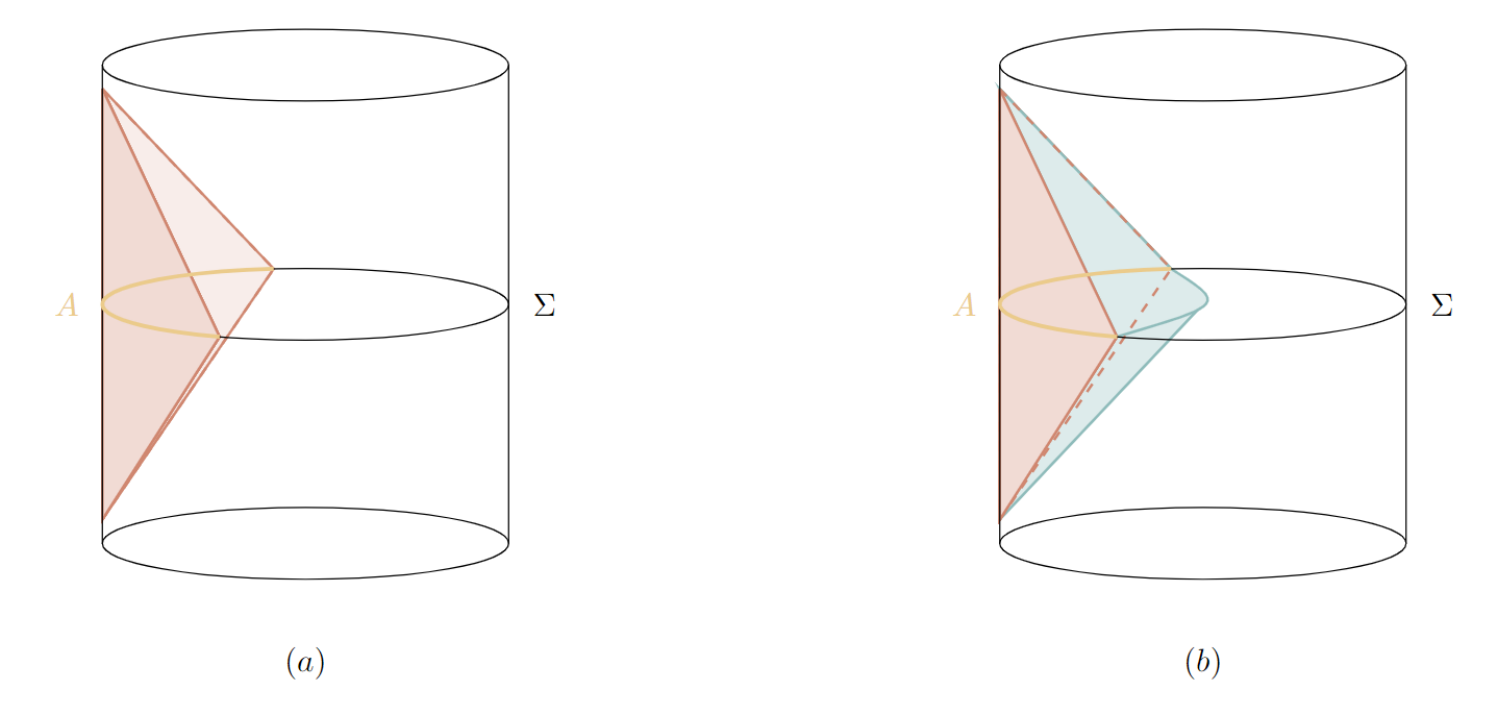}
    \caption{ (a) Domain of dependence of the spherical region of the boundary. (b) The entanglement wedge of the region A in the bulk which is called AdS-Rindler wedge. }
    \label{fig2}
\end{figure}

Consider the pure AdS$_{d+1}$ in the bulk. If we take the $t=0$ slice as the Cauchy surface and $A$ to be one hemisphere of $\Sigma$, the causal wedge of $A$ is the region of bulk that referred to as the AdS-Rindler wedge. Although it is naturally associated to a boundary region that covers half of the spatial surface, the patch can be mapped by an isometry to a patch that ends on arbitrary spatial region on the boundary.
The coordinate system that covers the AdS-Rindler patch is $(r, \tau, x)$ with the metric 
\begin{equation}
    ds^2 = -(r^2-1) d\tau ^2 + \frac{dr^2}{r^2-1} + r^2 dx^2
\end{equation}
where $x$ is the set of coordinates on the $(d-1)$ dimensional hyperbolic ball $H_{d-1}$. We can find the mode expansion of free scalar field in the AdS-Rindler wedge by solving the Klein-Gordon equation on this background
\begin{equation}\label{35}
    \phi(r, \tau, x)= \int \frac{d \omega}{2\pi} \frac{d\lambda}{2\pi} ~ \Big( f_{\omega\lambda}(r, \tau, x) b_{\omega\lambda}+f^*_{\omega\lambda}(r, \tau, x) b^\dagger_{\omega\lambda} \Big)
\end{equation}
where the modes $ b_{\omega\lambda}$ satisfy the usual commutation relation and the wave function is in the form of
\begin{equation} \label{50}
  f_{\omega\lambda}(r, \tau, x) = e^{-i\omega\tau} Y_{\lambda}(x) \psi_{\omega\lambda}(r).  
\end{equation}
The exact expression for the $ \psi_{\omega\lambda}(r)$ in terms of hypergeometric function is \cite{almheiri2015bulk}
\begin{multline}
     \psi_{\omega\lambda}(r)= M_{\omega\lambda} r^{-\Delta} (1-\frac{1}{r^2})^{-i\omega/2} 
     F \Big(-\frac{d-2}{4} +\frac{\Delta}{2} - \frac{i\omega}{2}+\frac{1}{2} \sqrt{\frac{(d-2)^2}{4}-\lambda}, \\
     -\frac{d-2}{4} +\frac{\Delta}{2} - \frac{i\omega}{2}-\frac{1}{2} \sqrt{\frac{(d-2)^2}{4}-\lambda}, \Delta - \frac{d-2}{2} , \frac{1}{r^2} \Big)
\end{multline}
 that 
 \begin{equation}
    M_{\omega\lambda}= \frac{1}{\sqrt{2|\omega|}} \frac{\Gamma(-\frac{d-2}{4} +\frac{\Delta}{2} + \frac{i\omega}{2}+\frac{1}{2} \sqrt{\frac{(d-2)^2}{4}-\lambda})~\Gamma(-\frac{d-2}{4} +\frac{\Delta}{2} + \frac{i\omega}{2}-\frac{1}{2} \sqrt{\frac{(d-2)^2}{4}-\lambda})}{\Gamma(\Delta - \frac{d-2}{2} )~\Gamma(i\omega)} .
 \end{equation}
By taking Fourier transformation of the boundary operator $ O(\tau, x)= \lim_{r \rightarrow \infty} r^\Delta \phi(r, \tau, x)$, we have
\begin{equation}\label{34}
    O_{\omega\lambda}= \int d\tau dx ~e^{i\omega\tau} Y_\lambda ^* (x) O(\tau, x)
\end{equation}
that is in the form of $O_{\omega\lambda}= M_{\omega\lambda}b_{\omega\lambda} $. Therefore, $ \hat{O}_{\omega\lambda}= \frac{1}{M_{\omega\lambda}} O_{\omega\lambda}$ is the boundary operator identified with Rindler mode functions
\begin{equation}\label{56}
   \hat{O}_{\omega\lambda}= b_{\omega\lambda} .
\end{equation}
By substituting (\ref{34}) into (\ref{35}) and exchange the order of integration  we get 
\begin{equation}
     \phi (r, \tau, x)=\int d\tau' dx'~ K(r, \tau, x|\tau', x') O(\tau', x')
\end{equation}
where the smearing function is 
\begin{equation}\label{26}
     K_R(r, \tau, x|\tau', x') = \int \frac{d \omega}{2\pi} \frac{d\lambda}{2\pi} ~ \frac{1}{M_{\omega\lambda}} f_{\omega\lambda} (r, \tau, x)~ e^{i\omega \tau'} Y^*_\lambda(x').
\end{equation}

The issue here is that if we substitute the exact expression of $f_{\omega\lambda} (r, \tau, x)$ in (\ref{26}), we find out that the integral does not converge for any choice of bulk and boundary points \cite{morrison2014boundary, leichenauer2013ads, rey2014scanning}.
In the original paper \cite{hamilton2006holographic}, authors argued that they can make the integral convergent by analytically continuation of $x$ coordinates. However, there is still this question that if it is actually well-defined in the physically correct Lorentz signature. The issue was illuminated in \cite{morrison2014boundary} when they gave an interpretation of the divergent smearing function in the context of distribution theory.

\section{Petz Theorem proof in finite dimensions}
\label{appendixpetz}

This appendix closely follows the discussion on \cite{petz2003monotonicity}. 

Consider a finite dimensional Hilbert space $\mathcal{H}$
with  the Hilbert-Schmidt inner product on the algebra of operators acting on  $\mathcal{H}$ as $\langle a,b \rangle = Tr (a^{\dagger}b)$ for all $ a,b \in L(\mathcal{H}) $. The action of a modular operator for all $a \in L(\mathcal{H}) $ is  $\Delta (a) = \sigma a \rho^{-1}$  where $\rho$ and $\sigma$ are positive definite Hermitian operators in $L(\mathcal{H})$. Consider another Hilbert space $L(\mathcal{K})$ and let $\mathcal{E} : L(\mathcal{}) \xrightarrow{} L(\mathcal{K})$ be a completely positive and trace preserving (CPTP) map, then $\mathcal{E}^{*}(1)=1$ for the dual of $\mathcal{E}$ with respect to the Hilbert-Schmidt inner product.   

Since $\mathcal{E}^{*}$ is also completely positive,
\begin{equation}
    \begin{pmatrix}
A & B \\
C & D 
\end{pmatrix} \geq 0 \implies   \begin{pmatrix}
\mathcal{E}^{*}(A) & \mathcal{E}^{*}(B) \\
\mathcal{E}^{*}(C) & \mathcal{E}^{*}(D) 
\end{pmatrix} \geq 0
\end{equation}
For some $a \in L(\mathcal{H}) $,
\begin{equation}
    \begin{pmatrix}
1 & a \\
0 & 0 
\end{pmatrix}^{*}  \begin{pmatrix}
1 & a \\
0 & 0 
\end{pmatrix} =   \begin{pmatrix}
1 & a \\
a^{*} & a^{*}a 
\end{pmatrix} \geq 0,
\end{equation}
then one has $\begin{pmatrix}
1 & \mathcal{E}^{*}(a) \\
\mathcal{E}^{*}(a^{*}) & \mathcal{E}^{*}(a^{*}a) 
\end{pmatrix} \geq 0$. For arbitrary $\eta$, one then has
\begin{equation}
    \langle \begin{pmatrix}
1 & \mathcal{E}^{*}(a) \\
\mathcal{E}^{*}(a^{*}) & \mathcal{E}^{*}(a^{*}a)
\end{pmatrix}\begin{pmatrix}
-\mathcal{E}^{*}(a) \eta \\
\eta
\end{pmatrix}, \begin{pmatrix}
-\mathcal{E}^{*}(a) \eta \\
\eta
\end{pmatrix} \rangle \geq 0.
\end{equation}
After a bit of algebra, one gets
\begin{equation}
    (\mathcal{E}^{*}(a^{*})\mathcal{E}^{*}(a)-\mathcal{E}^{*}(a^{*}a))\eta^{2} \leq 0
\end{equation}
which gives the Schwartz inequality for dual map $\mathcal{E}^{*}$, $\mathcal{E}^{*}(a^{*}a) \geq \mathcal{E}^{*}(a^{*})\mathcal{E}^{*}(a)$.

The relative entropy is defined as, 
\begin{equation}
    S(\rho||\sigma) = \tr (\rho (\log \;\rho - \log\; \sigma)) \qquad\text{ for } supp \; \rho \subseteq supp \; \sigma.
\end{equation}
otherwise it is defined to be infinite. One can see that $\Delta = \sigma \rho^{-1}$ and define also $\Delta_{0} = \mathcal{E}(\sigma) \mathcal{E}(\rho)^{-1}$.\footnote{Here and in what follows, we write $\Delta(1)$ and $\Delta_{0}(1)$ as $\Delta$ and $\Delta_{0}$ respectively. }
If both $\rho$ and $\sigma$ are invertible, one can write
\begin{equation}\label{eq 5.6}
\begin{split}
     S(\rho||\sigma) &= - \langle \rho^{1/2}, \log  \Delta \; \rho^{1/2}\rangle,\\
     S(\mathcal{E}(\rho)||\mathcal{E}(\sigma)) &= - \langle \mathcal{E}(\rho)^{1/2}, \log  \Delta_{0} \;\mathcal{E}(\rho)^{1/2}\rangle
\end{split}
\end{equation}
To relate the two equations above, we need some kind of relationship between $\rho$ and $\mathcal{E}(\rho)$ and the two modular operators. For some $x$, the norm of $x\mathcal{E}(\rho)^{1/2}$ is given by
\begin{equation}
\begin{split}
     ||x\mathcal{E}(\rho)^{1/2}||^{2} &= Tr(\mathcal{E}(\rho)^{1/2}x^{*}x\mathcal{E}(\rho)^{1/2})\\
                        &= Tr(\rho \mathcal{E}^{*}(x^{*}x)) \geq Tr(\rho \mathcal{E}^{*}(x)\mathcal{E}^{*}(x^{*})) 
\end{split}
\end{equation}
Thus one has $||x\mathcal{E}(\rho)^{1/2}||^{2} \geq ||\mathcal{E}^{*}(x)\rho^{1/2}||^{2}$. This means, one can define an operator $V$ such that 
\begin{equation}\label{eq 5.8}
    Vx\mathcal{E}(\rho)^{1/2}=\mathcal{E}^{*}(x)\rho^{1/2} \text{ and } V^{*}V\leq 1
\end{equation}
Considering \eqref{eq 5.8} for $x=1$ and squaring it, one gets $V\mathcal{E}(\rho)V^{*}=\rho$ and $V^{*}\rho^{-1}V = \mathcal{E}(\rho)^{-1}$. In addition, it also follows from the previous two formulas that $V^{*}\Delta V\leq \Delta_{0}$.

Coming back to the relative entropy in \eqref{eq 5.6}, a useful way to write logarithm is as the integral below
\begin{equation}
   - \log\; x = \int^{\infty}_{0} \Big((x+t)^{-1}-(1+t)^{-1}\Big)\; dt
\end{equation}
Thus,
one can rewrite the relative entropy as
\begin{equation}\label{eq 5.10}
   S(\rho||\sigma) = \int^{\infty}_{0}  \Big(\langle \rho^{1/2}, (\Delta + t)^{-1} \; \rho^{1/2}\rangle-(1+t)^{-1} \Big)\; dt
\end{equation}
similarly for $S(\mathcal{E}(\rho)||\mathcal{E}(\sigma))$. 

Using \eqref{eq 5.8} again for $x=1$, one can see that $\langle \rho^{1/2}, (\Delta + t)^{-1} \; \rho^{1/2}\rangle= \langle \mathcal{E}(\rho)^{1/2}, V^{*}(\Delta + t)^{-1} \; V\mathcal{E}(\rho)^{1/2}\rangle$. But one has the follow relation,
\begin{equation}\label{eq 5.11}
   V^{*}(\Delta + t)^{-1}V \geq (\Delta_{0} + t)^{-1}
\end{equation}
This is because $V^{*}\Delta V \leq \Delta_{0}$ and $(x+t)^{-1}$ is monotonically decreasing function. Thus one gets
\begin{equation}\label{eq 5.12}
  \langle \rho^{1/2}, (\Delta + t)^{-1} \; \rho^{1/2}\rangle \geq \langle \mathcal{E}(\rho)^{1/2}, (\Delta_{0} + t)^{-1} \; \mathcal{E}(\rho)^{1/2}\rangle.
\end{equation}
By substituting it in \eqref{eq 5.10}, this in turn implies that
\begin{equation}
  S(\rho||\sigma)\geq S(\mathcal{E}(\rho)||\mathcal{E}(\sigma))
\end{equation}
for any CPTP map $\mathcal{E}$. This inequality is known in the literature as Ulhmann's theorem of monotonicty of relative entropy. 

The interesting case is when the relative entropies are equal, which would imply equality in \eqref{eq 5.12} that is
\begin{equation}
  \langle \mathcal{E}(\rho)^{1/2}, V^{*}(\Delta + t)^{-1} \; V\mathcal{E}(\rho)^{1/2}\rangle = \langle \mathcal{E}(\rho)^{1/2}, (\Delta_{0} + t)^{-1} \; \mathcal{E}(\rho)^{1/2}\rangle
\end{equation}
where we used \eqref{eq 5.8} on the left side. Thus one has, $V^{*}(\Delta + t)^{-1} \; \rho^{1/2}=(\Delta_{0} + t)^{-1} \; \mathcal{E}(\rho)^{1/2}$. Repeating the analysis for another monotonically decreasing function $(x+t)^{-2}$, one gets
\begin{equation}\label{eq 5.15}
  V^{*}(\Delta + t)^{-2} \; \rho^{1/2}=(\Delta_{0} + t)^{-2} \; \mathcal{E}(\rho)^{1/2}
\end{equation}
It is then straightforward to show that $||V^{*}(\Delta + t)^{-1} \; \rho^{1/2}||^{2}= ||(\Delta + t)^{-1} \; \rho^{1/2}||^{2}$. For $V$ with property given in \eqref{eq 5.8}, one can show that 
\begin{equation}
\begin{split}
 (\Delta + t)^{-1} \; \rho^{1/2} &= VV^{*} (\Delta + t)^{-1} \; \rho^{1/2}\\
 &= V^{*} (\Delta_{0} + t)^{-1} \; \mathcal{E}(\rho)^{1/2}
\end{split}
\end{equation}
where in the second line we used \eqref{eq 5.8} and \eqref{eq 5.11}. Since the equality \eqref{eq 5.15} can be shown to hold for any same negative power of $\Delta + t$ and $\Delta_{0} + t$, the Stone-Wiestrass theorem implies that
\begin{equation}
Vf(\Delta_{0})\mathcal{E}(\rho)^{1/2}= f(\Delta)\rho^{1/2}
\end{equation}
for any continuous function $f$. Using \eqref{eq 5.8}, one can rewrite it as $\mathcal{E}^{*}(f(\Delta_{0}))=f(\Delta)$.

\subsection{Petz Theorem}

The theorem states that, for a CPTP map $\mathcal{E}$, $ S(\rho||\sigma) = S(\mathcal{E}(\rho)||\mathcal{E}(\sigma))$ if and only if there is another CPTP map $\mathcal{P}_{\sigma, \mathcal{E}}$ such that
\begin{equation}
\mathcal{P}_{\sigma, \mathcal{E}} \circ \mathcal{E}(\rho)=\rho, \text{ and } \mathcal{P}_{\sigma, \mathcal{E}}\circ \mathcal{E}(\sigma)=\sigma.
\end{equation}
\textbf{The proof:}

Assume there is some CPTP map $\mathcal{P}_{\sigma, \mathcal{E}}$ such that $\mathcal{P}_{\sigma, \mathcal{E}} \circ \mathcal{E}(\rho)=\rho$ and $\mathcal{P}_{\sigma, \mathcal{E}} \circ \mathcal{E}(\sigma)=\sigma$ then
\begin{equation}
\begin{split}
     S(\rho||\sigma)&\geq S(\mathcal{E}(\rho)||\mathcal{E}(\sigma)) \geq S(\mathcal{P}_{\sigma,\mathcal{E}} \circ \mathcal{E}(\rho)||\mathcal{P}_{\sigma, \mathcal{E}} \circ \mathcal{E}(\sigma)) 
\end{split}
\end{equation}
but since $S(\mathcal{P}_{\sigma, \mathcal{E}} \circ \mathcal{E}(\rho)||\mathcal{P}_{\sigma, \mathcal{E}} \circ \mathcal{E}(\sigma)) =  S(\rho||\sigma)$, the above equation implies that $S(\rho||\sigma) = S(\mathcal{E}(\rho)||\mathcal{E}(\sigma))$.

Assume $S(\rho||\sigma) = S(\mathcal{E}(\rho)||\mathcal{E}(\sigma))$ then for any continuous function $f$
\begin{equation}
    \mathcal{E}^{*}(f(\Delta_{0}))=f(\Delta).
\end{equation}
Consider $f(\Delta)= (\Delta^{*})^{-1/2}(\Delta)^{-1/2}=\sigma^{-1/2} \rho \sigma^{-1/2}$ and $f(\Delta_{0})=\mathcal{E}(\sigma)^{-1/2} \mathcal{E}(\rho) \mathcal{E}(\sigma)^{-1/2}$. Then, the above equality after a little algebra gives,
\begin{equation}
   \sigma^{1/2} \mathcal{E}^{*}\big(\mathcal{E}(\sigma)^{-1/2} \mathcal{E}(\rho) \mathcal{E}(\sigma)^{-1/2}\big)\sigma^{1/2} = \rho.
\end{equation}
Thus considering the CPTP map $\mathcal{P}_{\sigma, \mathcal{E}}(.) =\sigma^{1/2} \mathcal{E}^{*}(\mathcal{E}(\sigma)^{-1/2} (.) \mathcal{E}(\sigma)^{-1/2})\sigma^{1/2}$, the equality $\mathcal{P}_{\sigma, \mathcal{E}} \circ \mathcal{E}(\sigma) = \sigma $ immediately follows and $\mathcal{P}_{\sigma, \mathcal{E}} \circ \mathcal{E}(\rho)=\rho$ follows from the equality of relative entropy.$\blacksquare$ 

This map $\mathcal{P}_{\sigma, \mathcal{E}}$ is usually called the Petz recovery map.

\bibliographystyle{JHEP}
\bibliography{refs}

\end{document}